\documentclass[fleqn,usenatbib,useAMS]{mnras}

\usepackage{graphicx}	
\usepackage{amsmath}	
\usepackage{amssymb}	
\usepackage{multicol}        
\usepackage{bm}		
\usepackage{pdflscape}	
\usepackage{soul}	

\def\ga{\mathrel{\hbox{\rlap{\hbox{\lower4pt\hbox{$\sim$}}}\hbox{$>$}}}}
\def\la{\mathrel{\hbox{\rlap{\hbox{\lower4pt\hbox{$\sim$}}}\hbox{$<$}}}}

\def\msun{$M$\mbox{$_{\normalsize\odot}$}}

\def\mini{$M_{\rm init}$}

\def\mdot{$\dot{M}$}

\def\lbol{$L$\mbox{$_{\rm bol}$}}

\def\teff{$T_{\rm eff}$}

\def\logl{$\log (L/L_\odot)$}

\setstcolor{red}

\usepackage[T1]{fontenc}
\usepackage{ae,aecompl}

\usepackage{newtxtext,newtxmath}

\title[A new \mdot\ prescription for RSGs]{A new mass-loss rate prescription for red supergiants}

\author[E. R. Beasor et al.]{
Emma R. Beasor$^{1,2}\thanks{e-mail: embeasor@gmail.com}\thanks{Hubble Fellow}$,
Ben Davies$^{2}$,
Nathan Smith$^{3}$,
Jacco Th. van Loon$^{4}$,
\newauthor
Robert D. Gehrz$^{5}$
\& Donald F. Figer$^{6}$
\\
$^{1}$NSF's National Optical-Infrared Astronomy Research Laboratory, 950 N. Cherry Ave., Tucson, AZ 85719, USA\\
$^{2}$Astrophysics Research Institute, Liverpool John Moores University, 146 Brownlow Hill, Liverpool L3 5RF, UK\\
$^{3}$Steward Observatory, University of Arizona, 933 N. Cherry Ave., Tucson, AZ 85721, USA\\
$^{4}$Lennard-Jones Laboratories, Keele University, ST5 5BG, UK\\
$^{5}$Minnesota Institute for Astrophysics, School of Physics and Astronomy, 116 Church Street SE, University of Minnesota, Minneapolis, MN 55455, USA\\
$^{6}$Rochester Institute of Technology, 54 Memorial Drive, Rochester, NY 14623, USA
}
\date{Accepted XXX. Received YYY; in original form ZZZ}

\pubyear{2019}
\begin{document}
\label{firstpage}
\pagerange{\pageref{firstpage}--\pageref{lastpage}}
\maketitle

\begin{abstract}
Evolutionary models have shown the substantial effect that strong mass-loss rates (\mdot s) can have on the fate of massive stars. Red supergiant (RSG) mass-loss is poorly understood theoretically, and so stellar models rely on purely empirical \mdot-luminosity relations to calculate evolution. Empirical prescriptions usually scale with luminosity and effective temperature, but \mdot\ should also depend on the current mass and hence the surface gravity of the star, yielding more than one possible \mdot\ for the same position on the Hertzsprung-Russell diagram. One can solve this degeneracy by measuring \mdot\ for RSGs that reside in clusters, where age and initial mass (\mini) are known. In this paper we derive \mdot\ values and luminosities for RSGs in two clusters, NGC 2004 and RSGC1. Using newly derived \mini\ measurements, we combine the results with those of clusters with a range of ages and derive an \mini-dependent \mdot-prescription. When comparing this new prescription to the treatment of mass-loss currently implemented in evolutionary models, we find models drastically over-predict the total mass-loss, by up to a factor of 20. Importantly, the most massive RSGs experience the largest downward revision in their mass-loss rates, drastically changing the impact of wind mass-loss on their evolution. Our results suggest that for most initial masses of RSG progenitors, quiescent mass-loss during the RSG phase is not effective at removing a significant fraction of the H-envelope prior to core-collapse, and we discuss the implications of this for stellar evolution and observations of SNe and SN progenitors. \newline\newline\newline

\end{abstract}



\section{Introduction}

For the evolution of single stars, mass-loss prior to core collapse is arguably the most important factor affecting the evolution of a massive star across the Hertzsprung-Russel (HR) diagram, making it the key to understanding what mass-range of stars produce supernovae (SNe), and how these explosions will appear \citep{doggett1985comparative}. For initial masses below about 35 $M_{\odot}$, it is thought that most of the mass loss occurs during the red supergiant (RSG) phase, when strong winds dictate the onward evolutionary path of the star and potentially remove the entire H-rich envelope.

Uncertainty in the driving mechanism for RSG winds means mass-loss rate (\mdot) cannot yet be determined from first principles, and instead, stellar evolution models rely on empirical recipes to determine the outcome of their calculations \citep[e.g.][]{brott2011rotating,ekstrom2012grids,georgy2013grids,choi2016mist}. At present, the most commonly used \mdot-prescription comes from \citet{de1988mass}, a literature study in which many measurements of mass-loss were compiled. The sample sizes are small ($<$10 stars), highly heterogeneous in terms of mass and metallicity, have very uncertain distances from observations and analysis techniques that at best provide order-of-magnitude estimates compared to what is possible today. The relation itself contains large internal scatter ($\pm$ 0.5 dex), which could be the difference between a star losing its entire H-envelope, or almost none of it \citep[see e.g.][]{mauron2011mass}. This scatter has long been attributed to evolutionary effects \citep{van2009effects,van2005empirical,javadi2013uk}, with this more recently being confirmed by analysis of RSGs in clusters \citep[discussed later, ][]{beasor2016evolution,beasor2018evolution}. More modern efforts to update the RSG mass-loss rate prescription rely on samples which suffer from statistical biases, for example by selecting objects based on mid-IR brightness or circumstellar maser emission \citep{van2005empirical,goldman2017wind}, and hence are inevitably biased toward higher mass-loss rate objects. 

Uncertainties in RSG mass-loss in stellar models can have profound impact on evolutionary predictions \citep{smith2014mass}. Strong mass-loss during the RSG phase can cause the H-envelope to be peeled away, having direct consequences for predictions of SN rates \citep[e.g.][]{georgy2012yellow,smith11} and the Humphreys-Davidson (H-D) limit \citep{humphreys1979studies,davies2018humphreys}. Indeed, RSG mass loss has been suggested as a potential route to produce luminous blue variables (LBVs) or yellow hypergiants (YHGs) at masses lower than previously expected \citep[e.g.][]{groh2013massive}. High mass-loss rates during the RSG phase, particularly in the final $\sim$100s of years prior to core-collapse, are also invoked to explain the observational features of many Type II SNe, especially those of Type IIn \citep{smith09}. Slow rise times, bright initial peaks in the light curve, and narrow emission lines seen in the spectrum during the first few days after explosion are thought to be caused by the exploding star colliding with a dense layer of circumstellar material \citep[CSM,][]{chugai2004type}. 

\citet{beasor2016evolution,beasor2018evolution} have shown that the large dispersion on the \mdot-luminosity relation  \citep[e.g.][]{mauron2011mass} is vastly reduced when using RSGs within clusters as opposed to field stars. Using new age estimates for each cluster \citep{beasor2019ages}, in this paper we target RSGs in clusters again, further expanding the sample to include the younger cluster RSGC1 (where the RSGs are initially more massive) and older cluster NGC 2004 (where the RSGs are initially less massive) allowing us to probe how the \mdot-luminosity relation changes as a function of initial mass and age. Using this, we can create an initial mass-dependent \mdot-prescription and compare it to the current implementation of mass-loss in stellar models. 

In Secton \ref{section:observationsc4} we describe the sample of clusters and data used, in Section \ref{section:sedmodeling} we describe the dust shell models and fitting procedure, in Section \ref{section:resultsc4} we discuss the results and describe the method of determining \lbol, and finally in Section \ref{section:discussionc4} we discuss the findings in relation to other mass loss rate prescriptions, and consequences for stellar evolution.

\section{Observations}\label{section:observationsc4}
\subsection{Sample selection}\label{section:samplec4}
In our previous works we argue that the cause for large dispersion in many \mdot-prescriptions is due to the studies' use of field stars, where parameters such as initial mass and metallicity are unconstrained \citep{beasor2016evolution,beasor2018evolution}. For this reason, in the study presented here we focus solely on RSGs in clusters, for which initial mass and metallicty are constrained. We also require clusters that span a range of ages, in order to see how the \mdot-luminosity relation changes as a function of initial mass, ideally across the full range of RSG masses. The sample comprises five RSG rich clusters of varying ages, NGC 2100, NGC 7419, $\chi$ Per, RSGC1 and NGC 2004 (see Table \ref{table:clusterinfo} for cluster properties). By including a younger cluster in our sample, we will be able to anchor down the \mdot-luminosity relation for high-mass RSGs. As the He-burning lifetime for RSGs is very short, we can assume all of the RSGs currently in the RSG phase in a given coeval cluster are very similar in initial mass, to within $\sim$1\msun\  \citep{georgy2013grids}. Because of this, we will be able to derive an \mdot-luminosity relation dependent on the initial mass of the star. It can effectively be assumed that each RSG within a given cluster can be considered to be the same star at a different stage of evolution. 

Clusters NGC 2100, NGC 7419, $\chi$ Per and NGC 2004 have all been discussed in detail in previous papers \citet{beasor2016evolution,beasor2018evolution} and \citet{beasor2019ages}. 

\begin{table*}

\centering

\caption{Cluster properties.  }
\label{table:clusterinfo}
\begin{tabular}{lccccccc}

\hline
Cluster & Distance (kpc) & Age (Myr)&  Initial mass (\msun) & $A_{\rm V} (mag)$& $N_{\rm RSG}$ & References \\ [0.5ex] 
\hline
NGC 2100 & 50$\pm$0.1 & $21\pm1$ & $10\pm1$&0.5 & 19&1,2,5 \\
NGC 7419 &2.93$^{+0.32}_{-0.26}$& $20\pm1$ &$11\pm1$ &6.33$\pm$0.22 & 5&2,3 \\
$\chi$ Per & 2.25$^{+0.16}_{-0.14}$ &$21\pm1$& $11\pm1$&1.22$\pm$0.22&8 &2,3 \\
RSGC1 &6.6$\pm$0.9  &$7\pm2$ &$25\pm2$ & $25\pm2$$^\dagger$ & 15&4,6 \\
NGC 2004 & 50$\pm$0.1  &$23\pm1$ &$9\pm1$ &0.07 &7 &1,2\\

\hline

\end{tabular}

$^{1}$\citet{pietrzynski2013eclipsing}, $^{2}$\citet{beasor2019ages}, $^{3}$\citet{davies2019distances}\\
$^{4}$\citet{davies2008cool}, $^{5}$\citet{niederhofer2015no},$^{6}$\citet{figer2006discovery}\\
$^\dagger$ Converted from $A_{\rm K}$ using the extinction law of \citet{koornneef1983near}.

\end{table*}

\subsubsection{RSGC1}
First studied in \citet{figer2006discovery}, Galactic cluster RSGC1 was notable for its high number of RSGs. \citet{davies2008cool} estimated the age of RSGC1 by placing isochrones over the full range of RSGs in the cluster, for which they determined \teff\ and \lbol. The kinematic distance to the cluster was found to be 6.6 $\pm$ 0.9 kpc.

Unlike the other clusters in this sample, RSGC1 has high foreground extinction that is non-negligible in the mid-IR \citep[$A_{\rm k} = 2.74\pm0.2$ mag,][]{figer2006discovery}. Indeed, the extinction is high enough that for many of the RSGs in the cluster the mid-IR bump at 10$\mu$m used to trace mass-loss can disappear due to the foreground sillicate absorption being comparable to the object's intrinsic emission. For this reason, the extinction law has had to be carefully derived. To do this, we use an archival Spitzer/IRS \citep{werner2004spitzer,gehrz2007spitzer,houck2004irs} spectrum of F14\footnote{F14 was outside of the field of view for the data collected here.} This is the lowest luminosity RSG in the cluster, with no detectable IR excess \citep{davies2008cool}. Under the assumption that the star has no IR excess, the extinction law can be obtained by dividing the IRS spectrum through by an appropriate model atmosphere. See Appendix \ref{section:appendixA} for a full description. As we are assuming F14 has no circumstellar extinction, we take the value of reddening towards F14 as the foreground extinction towards the cluster (see Table \ref{table:clusterinfo}).  

\subsubsection{NGC 2004}
NGC 2004 is an LMC cluster containing seven RSGs, with their cluster membership confirmed by their radial velocities \citep[$\sim$300 km/s,][]{massey2003evolution}. By comparing the colour-magnitude diagram of this cluster to {\tt PARSEC} isochrones \citep{bressan2012parsec}, \citet{niederhofer2015no} estimate a reddening value of $E(B-V)$ = 0.23 mag. The age for NGC 2004 found in \citet{beasor2019ages}, 24 $\pm$2 Myr, is older than suggested by \citet{niederhofer2015no}, see \citeauthor{beasor2019ages} for more details.

\subsection{New observations and data reduction}
For RSGC1, we obtained new mid-IR photometry from SOFIA+FORCAST \citep{gehrz2009new,young2012early,herter2012first}. The data were taken in Cycle 5 using FORCAST (Prog ID 05 0064, PI Nathan Smith). The cluster was observed in 5.5$\mu$m, 7.7$\mu$m, 11.1$\mu$m, 25.3$\mu$m and 31.5$\mu$m filters to cover the emission from red stellar continuum and the warm circumstellar dust. In particular these wavelengths cover the infrared excess and 10$\mu$m silicate bump feature used to model the dust shells of the RSGs. The data was reduced using the SOFIA data pipeline {\tt FORCAST Redux}. The data-products we use are the Level 3 flux-calibrated data. We used IDL program {\tt starfinder}\footnote{http://www.bo.astro.it/StarFinder/} to extract photometry using point source function (PSF) fitting. PSFs for several isolated stars were combined using median averaging, from which we created our fiducial PSF. The PSF profile then underwent halo smoothing in the outer regions. To extract photometry, the threshold for star detection was defined as 5-sigma above background (all RSGs have much greater significance than this limit). The errors were assumed to be dominated by the variance in the sky. The photometry for RSGC1 is shown in Table \ref{table:RSGC1photom}.

For NGC 2004 we used archival data from several sources, including the Magellanic Clouds Photometric Survey \citep[MCPS,][]{zaritsky2004magellanic}, DENIS \citep{epchtein1994denis,cioni2000denis}, 2MASS \citep{skrutskie2006two2}, IRAC \citep{fazio2004irac} and WISE \citep{wright2010wide}.

\begin{table*}
\centering

\caption{Photometry for RSGC1 from SOFIA-FORCAST. All photometry is in Jy. }
\label{table:RSGC1photom}
\begin{tabular}{lcccccc}

\hline
ID &5.5$\mu$m &7.7$\mu$m &11.1$\mu$m& 25.3$\mu$m &  31.5$\mu$m \\ [0.5ex] 
\hline
F01&$ 6.88\pm 0.05$ &$ 5.33\pm 0.03$ &$15.07\pm 0.10$ &$12.86\pm 0.06$ &$10.99\pm 0.06$ &  \\
F02&$ 7.10\pm 0.05$ &$ 5.88\pm 0.03$ &$16.04\pm 0.10$ &$14.74\pm 0.07$ &$12.78\pm 0.08$ &  \\
F03&$ 4.08\pm 0.05$ &$ 4.44\pm 0.03$ &$ 9.59\pm 0.10$ &$ 8.07\pm 0.06$ &$ 6.93\pm 0.05$ &  \\
F06&$ 2.76\pm 0.05$ &$ 2.96\pm 0.02$ &$ 3.51\pm 0.10$ &$ 1.89\pm 0.05$ &$ 1.71\pm 0.05$ &  \\
F07&$ 2.70\pm 0.05$ &$ 2.42\pm 0.03$ &$ 2.81\pm 0.10$ &$ 1.28\pm 0.06$ &$ 1.09\pm 0.07$ &  \\
F09&$ 2.62\pm 0.05$ &$ 2.56\pm 0.02$ &$ 3.35\pm 0.10$ &$ 1.54\pm 0.06$ &$ 1.44\pm 0.05$ &  \\
F10&$ 2.06\pm 0.05$ &$ 2.10\pm 0.02$ &$ 3.03\pm 0.10$ &$ 1.80\pm 0.06$ &$ 1.81\pm 0.05$ &  \\
F12&$ 1.66\pm 0.05$ &$ 1.60\pm 0.03$ &$ 2.03\pm 0.10$ &$ 1.17\pm 0.06$ &$ 0.54\pm 0.06$ &  \\
F13&$ 4.30\pm 0.05$ &$ 3.18\pm 0.02$ &$ 7.28\pm 0.10$ &$ 8.12\pm 0.06$ &$ 8.51\pm 0.05$ &  \\

\hline
\end{tabular}
\end{table*}

\subsection{Determining cluster ages}
By studying RSGs in stellar clusters it is possible to determine ages and RSG initial massses ($M_{\rm ini}$) by fitting isochrones to observations. Many studies use the cluster main sequence turn off (MSTO) as an anchor point to determine the age. However as shown in \citet{beasor2019ages}, the presence of binary products (e.g. mergers or mass gainers) which appear brighter than the `true' MSTO, can cause the age of the cluster to be underestimated, and suggest RSG masses that are too high. For this reason, it was necessary to develop a new age diagnostic for star clusters, insensitive to the effects of rotation or binary evolution.

Here, we use the lowest luminosity RSG method to determine an age for the cluster, discussed at length in \citet{beasor2019ages}. This method relies upon the assumption that the lowest luminosity RSG is that which is least susceptible to the effects of binary interaction and rotation. The ages found from the lowest \lbol\ RSG are shown in Table \ref{table:clusterinfo}\footnote{Note that the ages presented in this paper supersede the results from \citet{beasor2016evolution,beasor2018evolution}. Previously, for NGC 7419 and $\chi$ Per, the ages were estimated by comparing isochrones to the MSTO of the cluster \citep{marco2013ngc,currie2010stellar}, a method which may have been affected by the presence of blue straggler-like objects.}. The RSGs in this  sample span initial masses between 9 and 25\msun, covering the majority of the initial mass range expected to end their lives as Type II-P SNe \citep[e.g.][]{meynet2003stellar}\footnote{ For comparison, the ages determined for each cluster using the MSTO yielded cluster ages that were younger by an average of 11Myr and hence implied higher RSG masses. See Table 2 in \citet{beasor2019ages} for details. }. 

\section{Spectral energy distribution modeling}\label{section:sedmodeling}
The model setup has been described in detail in \citet{beasor2016evolution} and again in \citet{beasor2018evolution}. Below we will briefly describe the model setup and chosen input parameters.

Throughout this work we use dust shell models from {\tt DUSTY} \citep{ivezic1999dusty}, a code which solves the radiative transfer equation for a star surrounded by a spherically symmetric layer of dust of a given optical depth ($\tau_{\rm V}$, optical depth at 0.55$\mu$m), inner dust temperature ($T_{\rm in}$) at the innermost radius of the dust shell ($R_{\rm in}$) and radial density profile ($\rho_{\rm r}$).  

Dust surrounding a star leaves signatures in the output spectrum, as the light is absorbed and re-processed. From this it is possible to determine the chemical composition of the dust surrounding the star, and how much of it there is. The 10$\mu$m silicon `bump', indicative of oxygen rich dust, has been observed around many RSGs \citep[e.g.][]{ohnaka2008spatially}, and hence we opted for silicate dust as described by \citet{draine1984optical} with a fiducial grain size ($a$) with a radius of 0.3$\mu$m\footnote{In \citet[][]{beasor2016evolution} it was shown that varying the grain size had no effect on the derived mass-loss rate to within the errors, see paper for a detailed discussion.}. We assume a gas-to-dust ratio ($r_{\rm gd}$) of 200 for the MW clusters and 500 for the LMC cluster \citep{van2005empirical}. (Note that these differences in gas to dust ratios account for how metallicity influences rates derived from observations, but it does not account for any metallicity dependence in the driving mechanism of the wind.)  For all stars we assumed a grain bulk density $\rho_{\rm d}$ of 3 g cm$^{-3}$. Together, these parameters allow a dust shell mass to be derived for each model. 

To calculate \mdot, we also need to make assumptions about the density profile of the dust and the outflow velocity of the winds. As in \citet{beasor2016evolution,beasor2018evolution}, we have used a steady state wind with a density distribution that falls off with $r^{-2}$. The stars in this  sample do not have measured outflow velocities, we therefore use a uniform speed of 25$\pm$5 km/s, consistent with measurements taken for other RSGs \citep{richards1998maser,van2001circumstellar}\footnote{It should be noted that there is evidence that RSG wind speed correlates with metallicity \citep[e.g.][]{marshall2004asymptotic,goldman2017wind}. The effect of this on \mdot\ is discussed in \citet{beasor2018evolution}.}. It is possible that the more massive RSGs will have faster wind speeds than the less massive RSGs due to the more massive objects having higher surface gravities. If this were the case, we would systematically underestimate \mdot\ for the most massive RSGs in our sample, but it is likely this effect would be minimal. With this, we can calculate \mdot\ using the following equation

 \begin{equation} \dot{M} = \frac{16\pi}{3} \frac{R_{in} \tau_{V}  \rho_d a v_\infty}{Q_{V}}r_{\rm gd}
 \end{equation}

\noindent where $Q_V$ is the extinction efficiency of the dust \citep[as defined by the dust grain composition,][]{draine1984optical}.      

It is also necessary to assume an effective temperature ($T_{\rm eff}$) for the RSGs, as $T_{\rm eff}$ defines the input spectral energy distribution that will be reprocessed by the surrounding dust shell. There is some controversy over the temperatures of RSGs \citep{levesque2005effective,davies2013temperatures}, and so this study explores a temperature range of 3600-4200K, with 3900K being the fiducial effective temperature. 
 In this work we have used a grid spanning inner dust temperatures of 100 - 1200K in steps of 100K and optical depth values of 0 - 4 in steps of $\sim$ 0.08. For each {\tt DUSTY} model we compute synthetic photometry by interpolating the model flux onto the filter profiles. We then use $\chi^2$ minimisation to find the best fit model as in the following equation
 
\begin{equation}
\chi^2 =\sum_i \frac{ (O_{i}-E_{i})^2 }{\sigma_i^2}
\end{equation}
where $O$ is the observed photometry, $E$ is the model photometry, $\sigma^{2}$ is the error and $i$ denotes the filter. In this case, the model photometry provides the ``expected'' data points. The best fitting model is that which produced the lowest  ${\chi^2}$. The ``error models" are the models that fit within the minimum $\chi^2$+10 limit. This limit was chosen so that the stars with the lowest measured \mdot, which were clearly consistent with non-detections, would have \mdot\ values consistent with 0 (or upper limits only). As our methodology is dominated by systematic effects (e.g. SED temperature, the shape of the extinction law), the assumption of purely Gaussian errors is invalid. It is for this reason that we do not use the formal limit for a 1$\sigma$ error, and instead define our error limit as the minimum $\chi^2$+10.


\section{Results}\label{section:resultsc4}
\begin{figure*}
  \caption{ \textit{Left panel:}  Model plot for F01 in RSGC1 including all contributions to spectrum. The silicate bump at 10$\mu$m is clearly visible suggesting a large amount of circumstellar material. \textit{Right panel:} Contour plot showing the degeneracy between $\chi^2$ values and best fitting \mdot\ values. The thickened contour highlights the models within the minimum $\chi^2$+10 limit. }
  \centering
  \label{fig:allcontc4}
     \includegraphics[height=7.0cm]{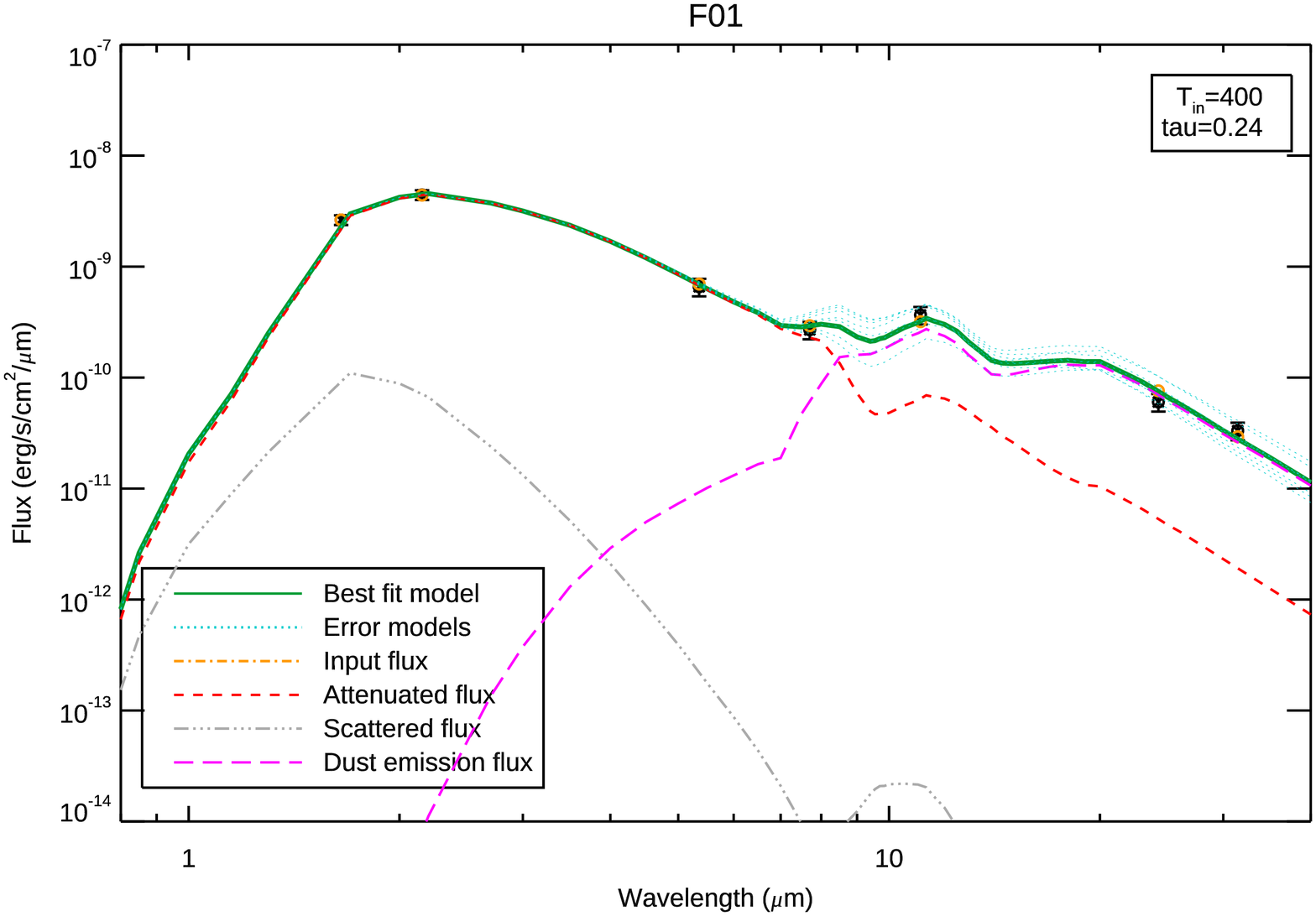}
     \includegraphics[height=7.0cm]{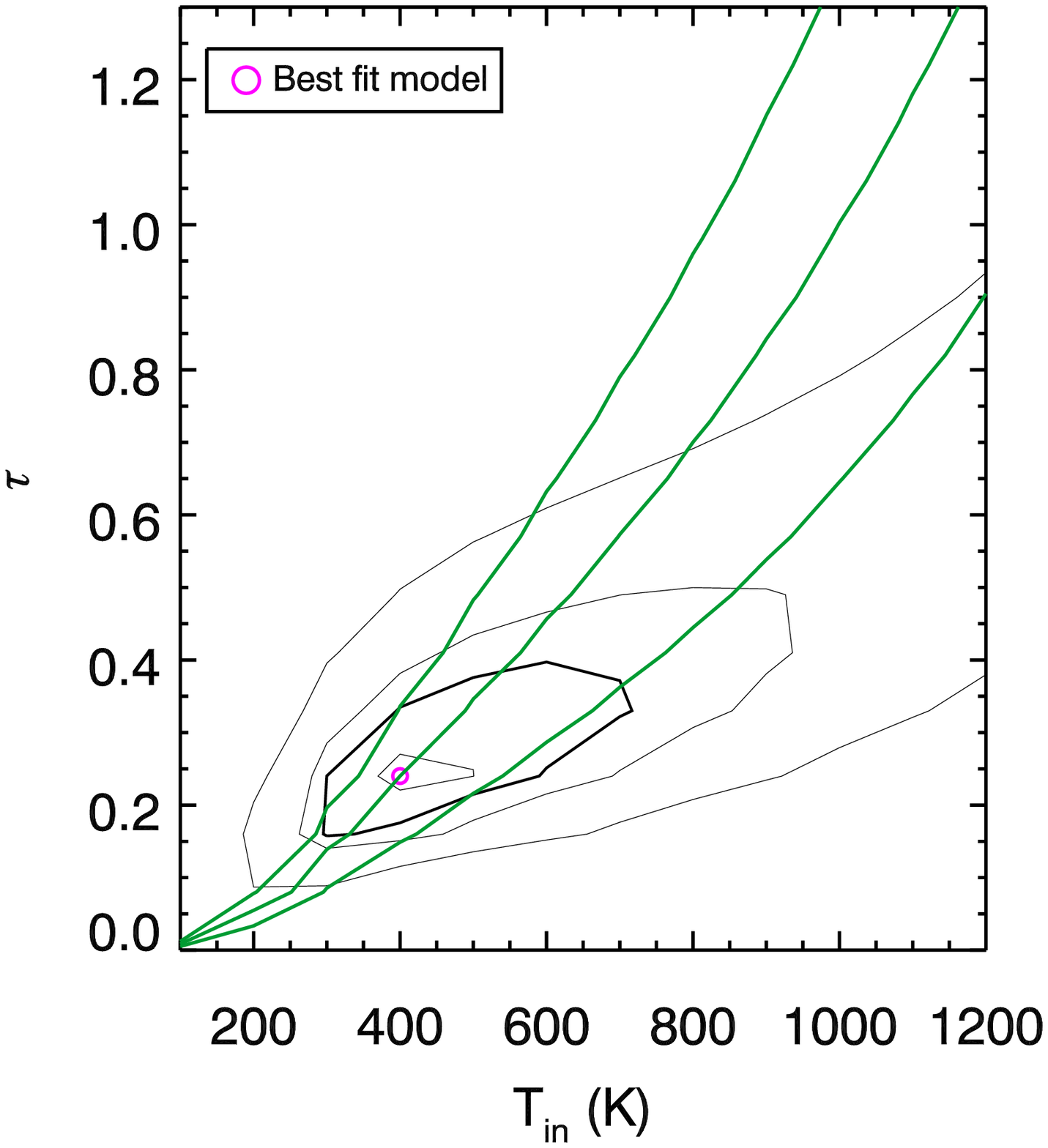}
\end{figure*}

The mass-loss rates and luminosities for both clusters are shown in Table \ref{table:resultsc4}. Figure \ref{fig:allcontc4} shows the best fit model for the brightest RSG in the sample, F01. The left panel of the plot shows the best fit model spectrum (green line), the models within the error range (blue dotted lines) as well as the photometric points, where the black crosses show the real photometry and orange circles show the model photometry. This plot also shows all contributions to the output spectrum, including the dust emission flux and flux from scattered light. The right hand panel shows the best fit model located on a $T_{\rm in}$ - $\tau$ plane with the mass-loss rate isocontours overplotted, demonstrating the degeneracy between $T_{\rm in}$ and $\tau_V$. 

Figure \ref{fig:mdotcompc4} shows $L_{\rm bol}$ versus \mdot\ for the two clusters presented here, from which we can see an increase in \mdot\ with luminosity. We have also included the results from clusters in previous papers \citet{beasor2016evolution,beasor2018evolution} corrected for new distances and ages. 

\subsection{Luminosities}
The luminosities for the RSGs in NGC 2004 were calculated by integrating under the observed spectral energy distribution (SED), as in \citet{davies2018initial}. We took all of the available photometry and integrated underneath the points using IDL routine {\tt int$\_$tabulated}\footnote{https://www.harrisgeospatial.com/docs/INT$\_$TABULATED.html}.  To include any flux that may be missing from shorter wavelengths, the SED was extrapolated using a blackbody curve that was fitted to the shortest wavelength available photometry, in this case $B$-band. Although it should be noted that the contribution to the overall luminosity from this region of the SED is extremely small ($<$0.01 dex). 

For RSGC1, we did not estimate the luminosity from the SED. This is because the shortest wavelength photometry available was at 2MASS-J, and the extrapolated flux would contribute a large fraction to the luminosity estimate. For this reason we use the best fit model from {\tt DUSTY} to extrapolate the fluxes below 1$\mu$m. Therefore, the errors are dominated by the uncertainty in $T_{\rm eff}$ and $A_{\rm V}$. 

The star F13 is anomalously red compared to the other RSGs in the cluster \citep{davies2008cool}, either due to circumstellar extinction or additional foreground extinction. It is therefore likely \lbol\ will be underestimated as we have assumed the same extinction value for all stars. When taking into account the extra extinction ($\Delta$$A_{\rm K}\sim 0.9mag$\footnote{This extinction corresponds to an $A_{\rm V}$ of 9 mag. If this extinction was due to CSM it would imply an extreme \mdot, which itself is not consistent with the modest mid-IR excess observed. This extra extinction is therefore likely foreground.}) the luminosity increases to \logl=5.39 (from \logl=5.18). Due to the uncertainty in the true luminosity of this star we have not included it in calculating an \mdot\ - luminosity relation for the cluster (see Section 6.1), though when adopting the higher extinction value for this star we note that it agrees perfectly with the other stars in the cluster. 

For NGC 7419 and $\chi$ Per, due to updated distances from Gaia \citep[][]{davies2019distances,gaiadr2}  the luminosities have also changed since they were published in \citet{beasor2018evolution}, and are now lower by an average of 0.1 dex. The \mdot\ values plotted are scaled in accordance with the updated luminosities.

\begin{table*}

\centering

\caption{Fitting results for the RSGs in RSGC1 and NGC 2004. $A_{\rm V}$ is the extinction due to the circumstellar wind at 0.55$\micron$.}
\label{table:resultsc4}
\begin{tabular}{llccccc}

\hline
Cluster & Star & $T_{\rm in}$ (K) & $\tau_V$ &  \mdot\ (10$^{-6}$\msun\ yr$^{-1}$) &  $L_{\rm bol}$  & $A_{\rm V}$ (mag) \\ [0.5ex] 
\hline


RSGC1 &F01&$ 400^{+ 300}_{- 100}$&$0.24^{+0.09}_{-0.08}$&$ 5.57^{+ 2.37}_{- 2.17}$&$ 5.58\pm 0.18$ & 0.03 \\
&F02&$ 500^{+ 200}_{- 200}$&$0.33^{+0.16}_{-0.09}$&$ 5.18^{+ 2.72}_{- 1.75}$&$ 5.56\pm 0.18$ & 0.05\\
&F03&$ 400^{+ 100}_{- 100}$&$0.24^{+0.17}_{-0.00}$&$ 4.18^{+ 3.08}_{- 0.84}$&$ 5.33\pm 0.08$ & 0.05\\
&F06&$ 600^{+ 600}_{- 200}$&$0.08^{+0.08}_{-0.00}$&$ 0.68^{+ 0.69}_{- 0.14}$&$ 5.32\pm 0.18$ & 0.01\\
&F07&$1000^{+ 200}_{- 400}$&$0.08^{+0.08}_{-0.00}$&$ 0.28^{+ 0.29}_{- 0.06}$&$ 5.31\pm 0.18$ & 0.01\\
&F09&$ 700^{+ 500}_{- 200}$&$0.08^{+0.08}_{-0.00}$&$ 0.52^{+ 0.53}_{- 0.10}$&$ 5.30\pm 0.18$ & 0.01\\
&F10&$ 500^{+ 500}_{- 100}$&$0.08^{+0.08}_{-0.00}$&$ 0.87^{+ 0.89}_{- 0.17}$&$ 5.28\pm 0.18$ & 0.01\\
&F12&$1200_{- 400}$&<0.08&$ 0.18^{+ 0.04}_{- 0.04}$&$ 5.22\pm 0.19$ & 0.03\\


\hline
NGC 2004 &SV* HV 2595&$1200_{- 200}$&$1.39^{+0.81}_{-0.57}$&$43.29^{+28.71}_{- 9.74}$&$ 5.15\pm 0.04$ & 0.98\\
&LHA 120-S 43&$-$&$-$&$ <1.09$&$ 4.85\pm 0.05$& - \\
&Cl* NGC 2004 E 33&$-$&$-$&$ <0.84$&$ 4.35\pm 0.05$&-\\
&W61 22-9&$ -$&$-$&$ <0.54$&$ 4.55\pm 0.05$&- \\
&Cl* NGC 2004 BBBC 431&$-$&$-$&$ <0.55$&$ 4.55\pm 0.05$ &-\\
&W61 18-13&$-$&$-$&$ <1.96 $&$ 4.58\pm 0.05$&- \\

\end{tabular}
\end{table*}

\begin{figure}
\centering
  \caption{ {\it Top panel:} \mdot\ versus $L_{\rm bol}$ for all clusters studied here. The dashed lines show the individual fits to each relation, shown in Table \ref{table:mdotrelations}. {\it Bottom panel: }Same as above, solid lines show fits to \mdot-\lbol\ relation once the gradient has been fixed. }
  \centering
  \label{fig:mdotcompc4}
    \includegraphics[width=\columnwidth]{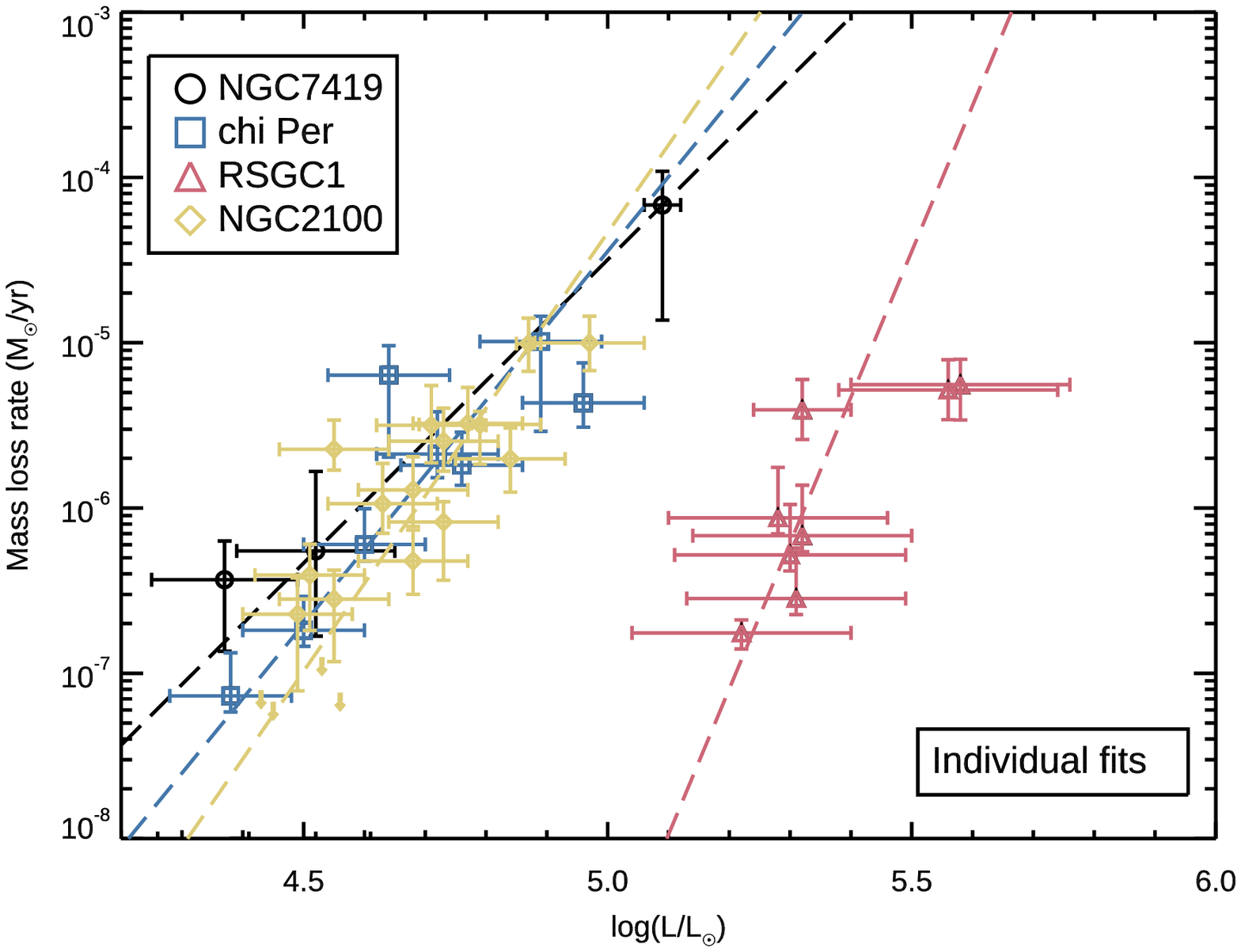}
      \includegraphics[width=\columnwidth]{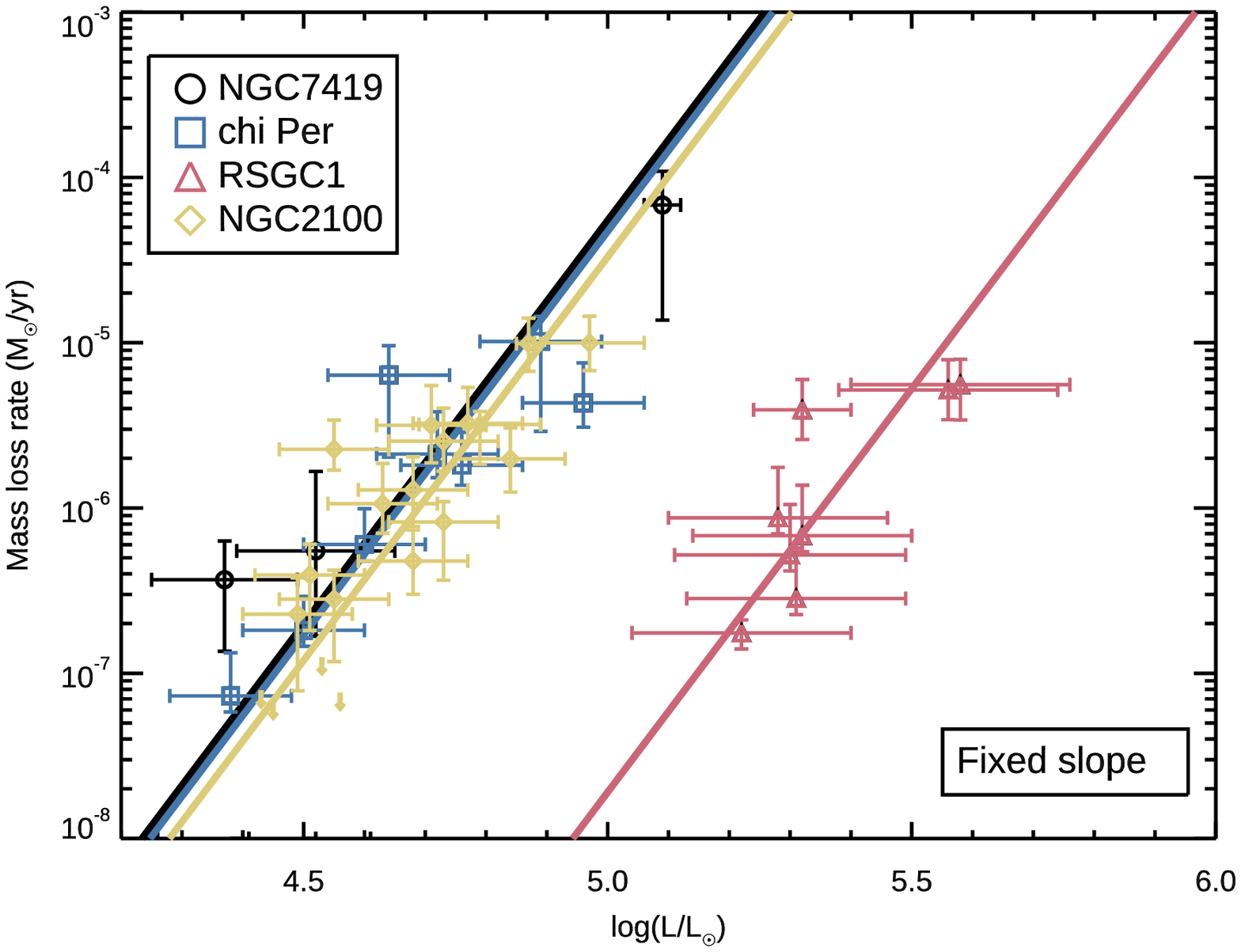}
\end{figure}

Figure \ref{fig:mdotcompc4} shows \mdot\ versus luminosity for four of the clusters presented in this work. For NGC 2004, only the most luminous star has an \mdot\ measurement; the rest of the stars in this sample are upper limits only. We therefore choose not to include these objects in any further analysis. 
 
\section{Discussion}\label{section:discussionc4}
\subsection{The \mdot-luminosity relation for red supergiants}
Empirically derived \mdot - prescriptions are vital input for stellar evolutionary models.  It is from the mass loss that the onward evolution of RSGs is predicted, as the amount of mass lost determines where the star ends up on the HR diagram, which in turn determines the final fate of the star. The most commonly used prescription, that of \citet{de1988mass}, was determined by compiling \mdot\ values for 271 field stars from various other studies. Of this sample, there are 15 RSGs included in the sample, with no constraints on initial mass. This prescription is dependent only on the luminosity of the star. 

We have previously shown that by keeping $M_{\rm ini}$ constrained, the \mdot - luminosity relation is a tighter correlation with a dispersion of only 0.4 dex \citep{beasor2016evolution,beasor2018evolution}. We now focus on different mass RSGs, including the higher mass RSGs in RSGC1, where the impact of mass-loss could be more significant. We cannot derive a relation for the RSGs in NGC 2004 as apart from the brightest star (SV* HV 2595) all of the measurements on \mdot\ are upper limits. We now use {\tt IDL} routine {\tt FITEXY}\footnote{https://idlastro.gsfc.nasa.gov/ftp/pro/math/fitexy.pro} to determine the \mdot-luminosity relations for all other clusters in the sample. From this we find a relation of 

\begin{equation}
\log(\dot{M} / M_\odot {\rm yr}^{-1} )= a + b\log(L_{\rm bol} / L_\odot)
\end{equation}

\noindent where the values of $a$ and $b$ are shown in Table \ref{table:mdotrelations} and are specific to each cluster.

\begin{table}
    \centering
        \caption{\mdot\ relation parameters for each cluster. The \mdot-luminosity relation is in the form $\log(\dot{M} / M_\odot yr^{-1} )= a + b\log(L_{\rm bol} / L_\odot)$. We also show the Pearson correlation coefficients for each relation. }
    \begin{tabular}{lccc}
    \hline
        Cluster & Offset ($a$) & Gradient ($b$) & Correlation Coeff.  \\
        \hline
         NGC 2100 & $-30.9\pm3.6$ & $5.3\pm0.8$& 0.79\\ 
         NGC 7419 & $-22.9\pm4.9$&$3.6\pm1.7$&0.97\\
         $\chi$ Per & $-27.0\pm4.9$&$4.5\pm1.0$&0.66\\
         RSGC1 & $-52.0\pm 51.2$ & $8.8\pm9.5$&0.87 \\
         
         \hline
    \end{tabular}

    \label{table:mdotrelations}
\end{table}

\begin{figure}
    \centering
    \caption{Initial mass versus \mdot-\lbol\ relation offset for each cluster. }
    \includegraphics[width=\columnwidth]{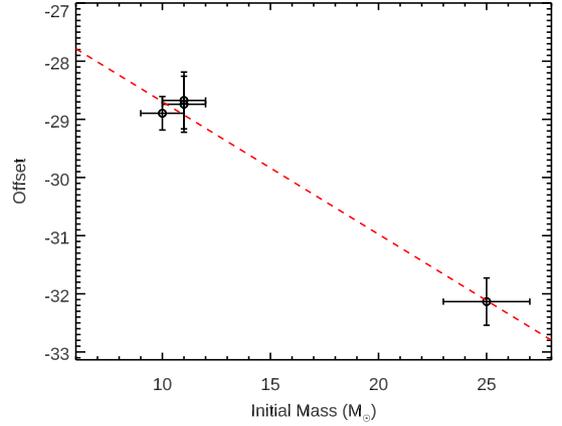}

    \label{fig:offsets}
\end{figure}

We now have \mdot-luminosity relations for RSGs across a range of initial masses. Using the updated cluster ages found in \citet{beasor2019ages} we have re-derived initial masses for RSGs, shown in Table \ref{table:clusterinfo}. All of the \mdot-luminosity relations are shown in Fig. \ref{fig:mdotcompc4}. The gradients of each \mdot-luminosity relation are consistent to within the errors. Taking the average of these values, we now fix the gradient of the \mdot-luminosity relation for each cluster, see the bottom panel of Fig. \ref{fig:mdotcompc4}. We choose to fix the gradient in order to reduce the number of degrees of freedom in the fit. By fixing the gradients, there is only one free parameter that needs to be calibrated, which in turn leads to more reliable results when extrapolating outside of the observed parameter space. From this, we find the mass-dependent offset. Figure \ref{fig:offsets} shows the relation of initial mass with offset. We can see that RSGC1 has a very different offset compared to the other clusters, which we interpret as a mass dependency of `b'. In the absence of data points in between the lower \mini\ clusters and RSGC1, we perform a simple linear fit. We discuss the potential implications of this limited sampling in \mini\ further in Section \ref{section:caveats}.  

With the \mdot-luminosity relation for each cluster, in combination with estimates of the initial masses of the RSGs in the clusters, we can now parameterise \mdot\ in terms of both \lbol\ and initial mass. A more general mass dependent \mdot-luminosity relation can be derived, 
\begin{equation}
\log(\dot{M} / M_\odot {\rm yr}^{-1} )= (-26.4-0.23\times M_{\rm ini} / M_{\odot} )+ b\log(L_{\rm bol} / L_\odot)
\end{equation}
 where $b = 4.8\pm 0.6$. This dependence of offset on initial mass explains why many other \mdot\ prescriptions have such high dispersions, as changing $M_{\rm ini}$ causes the relation to become `smeared' across luminosities. At fixed luminosity, RSGs have higher \mdot\ at lower initial mass. This is to be expected, since lower mass implies lower surface gravity, which presumably makes winds easier to drive. This is very important.  Not including the effects of stellar mass in past prescriptions, but extrapolating mass-loss prescriptions to very high luminosity, has caused stellar evolution codes to severely overestimate the influence of winds for the highest-mass RSGs.

 \subsection{Comparison to other \mdot-prescriptions}
\begin{figure*}
     \centering
     \includegraphics[width=18cm]{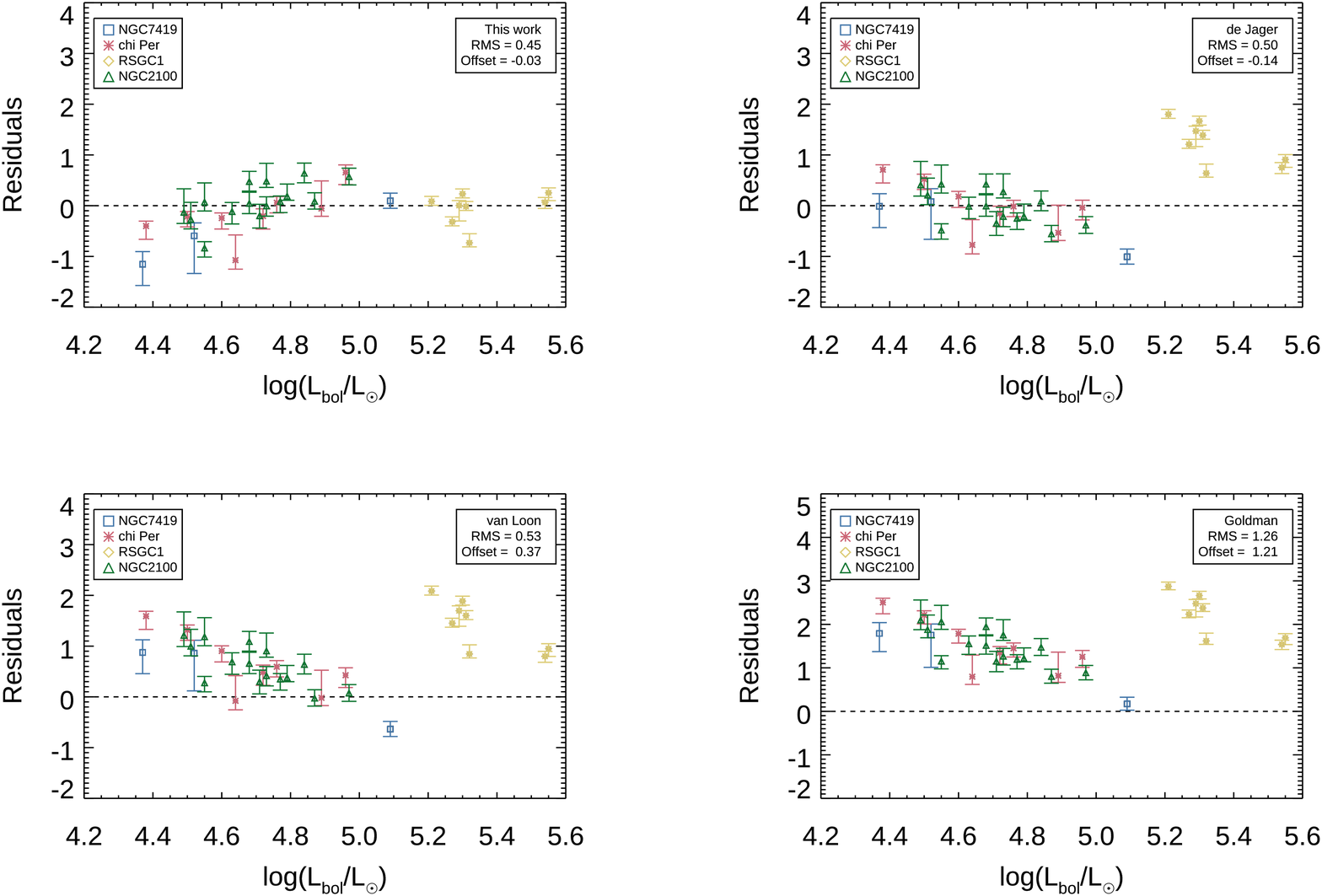}
     \caption{Residual \mdot\ values, defined as log(\mdot$_{\rm measured}$) - log(\mdot$_{\rm prescription}$), for each star using the \mdot-prescriptions from this work, \citet{de1988mass}, \citet{van2005empirical} and \citet{goldman2017wind}.}
     \label{fig:residuals}
 \end{figure*}

 We now compare the performance of our prescription to others commonly used in stellar evolutionary codes. To do this, we calculate the residuals for each prescription, by subtracting the mass-loss rate found from the relation to the measured value of \mdot. For comparison we compare the results to the de Jager prescription \citep{de1988mass}, van Loon \citep{van2005empirical} and the more recent \citet{goldman2017wind} prescription.  Results are shown in Fig. \ref{fig:residuals}, with the root mean square (RMS) and mean values shown in Table \ref{table:mdotrelations}. To estimate RMS and offset we use only mass-loss rates higher than 10$^{-6}\msun yr^{-1}$, as below this the value of \mdot\ is negligible.

%
              
Our prescription provides the most accurate and precise results, with an RMS of $\pm$0.45 dex. The dispersion on the \citet{de1988mass} prescription is larger ($\pm$0.50 dex) and in addition has a systematic overestimate of 0.12 dex. The offset becomes more significant at high luminosities. For the RSGC1 stars, the most luminous objects in our sample, dJ88 systematically overestimates \mdot\ by a factor of 1.3 dex. As illustrated in Fig. \ref{fig:residuals}, this is particularly evident for the highest luminosity stars (\logl\ $>$ 5), where the mass-loss rates are systematically overestimated by a factor of 10. The dJ88 prescription performs particularly badly for the highest \lbol\ (and hence initial mass) RSGs, for which \mdot\ presumably has the greatest potential effect.  

The van Loon and Goldman prescriptions both lead to large dispersions ($\pm$ 0.53 dex and $\pm$ 1.26 dex respectively) and in all cases over predict the amount of mass lost, by factors of $\times$2 and $\times$16 respectively (see Fig. \ref{fig:residuals}). As discussed in previous papers \citep{mauron2011mass,beasor2016evolution, beasor2018evolution} both studies select stars with enhanced mass-loss, by either selecting dust enshrouded objects \citep{van2005empirical} or maser emitters \citep{goldman2017wind}. It is likely that the stars chosen in these studies, are at the later stages of evolution and are experiencing the highest levels of mass loss, and hence are not representative for RSGs in the earlier phases of evolution. In their paper, \citet{van2005empirical} compared the \mdot\ values predicted by their prescription to measured \mdot\ values for Galactic RSGs, finding only the most extreme objects (e.g. VY CMa, VX Sgr) were consistent to within the errors. Looking at the results (bottom two panels in Fig. \ref{fig:mdotcompc4}), if one is to follow the residuals for a cluster, the dispersion at later stages of evolution (higher luminosities) is smaller, supporting the hypothesis that both the van Loon and Goldman prescriptions are applicable for RSGs at the end of their lives. While these prescriptions are perhaps not appropriate for input into stellar evolutionary models, they have the advantage of not requiring an initial mass, and so have the potential to be used to estimate \mdot\ for stars with strong pulsations (e.g. Mira variables).

\subsection{Total mass lost during the RSG phase}
How much mass is lost by a star prior to explosion is an important factor on the appearance of the eventual SN. It is predicted that stars with initial masses between 8 and 25\msun\ will evolve through the RSG phase before exploding as a Type II-P SN, while stars above this mass range are predicted to shed their outer envelope and explode in the blue region of the HR diagram.

\begin{figure}
    \centering
    \includegraphics[width=\columnwidth]{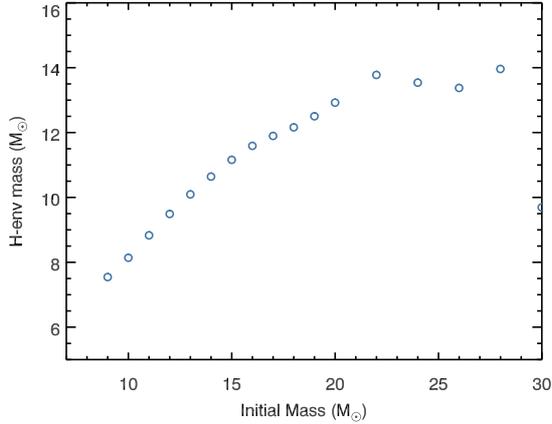}
    \caption{Mass of the H-rich envelope at the end of the MS for a star as a function of initial mass using the MIST mass tracks (see text for details). }
    \label{fig:MISTenvmass}
\end{figure}

There is a maximum limit to how much mass an RSG can lose, determined by the mass of the H-rich envelope. If this is removed completely, the star cannot remain in the red of the HR diagram, and instead will evolve back to the blue. Using the MIST models \citep{dotter2016mesa,choi2016mist,paxton2010modules,paxton2013modules,paxton2015modules}, it is possible to determine how envelope mass changes as a function of initial mass, see Fig. \ref{fig:MISTenvmass}. This figure shows the envelope mass for a star of a given initial mass at the beginning of the RSG phase, where envelope mass is estimated by subtracting the helium core mass from the mass of the star at the end of the MS. For an RSG with an initial mass of 20\msun\ to evolve to the blue of the HR diagram, it would have to lose $\sim$13\msun\ of mass during the RSG phase prior to explosion. If we assume the RSG phase is 10$^{6}$yrs, this would require an average sustained  \mdot\ of 10$^{-5}$\msun\ yr$^{-1}$, a mass-loss rate only observed for the brightest and most evolved RSGs in this sample.

\begin{figure}
	\centering
  	\caption{\mdot\ as a function of time using the Geneva mass tracks at 12, 15, 20 and 25\msun. At each timestep, we use the new \mdot-prescription derived here and calculate a new value for mass loss.  }
  	\centering
  	\label{fig:mdot_geneva}
			    \includegraphics[width=\columnwidth]{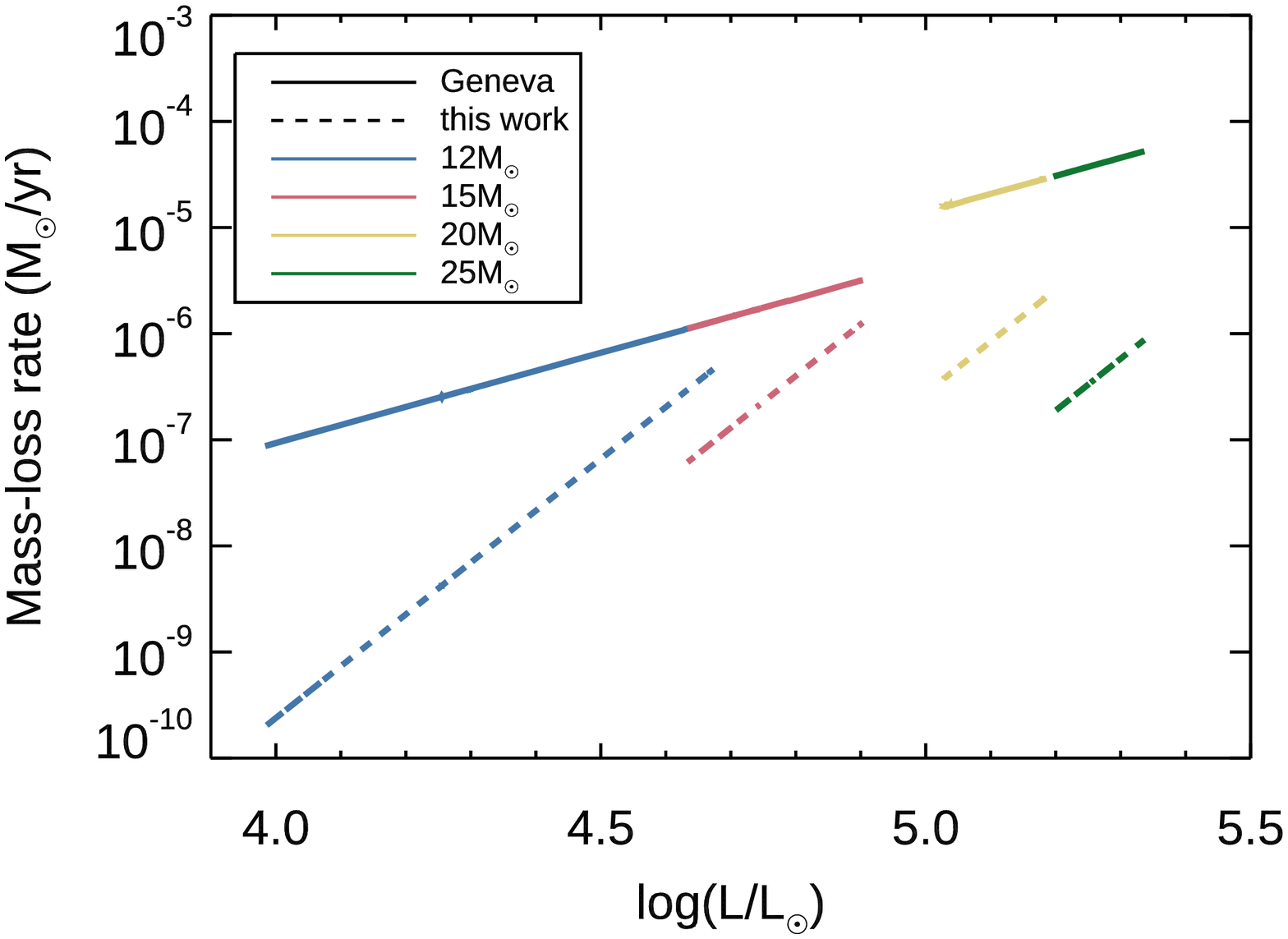}
	
\end{figure}

We now compare the amount of mass lost for 12, 15, 20 and 25\msun\ stars in the Geneva mass tracks\footnote{For the purposes of this study we compare thew new \mdot-prescription to Geneva models only as these models are optimised for massive stars in terms of how they are calibrated (for example overshooting and rotation). As well as this, they are also the most commonly used stellar evolutionary model in the field.}. For each mass track, we begin by plotting \mdot\ as a function of luminosity, shown in Fig. \ref{fig:mdot_geneva}. Note the increase in \mdot\ by a factor of 3 at masses of 20\msun\ and over; this arbitrary increase of \mdot\ is implemented in the models when the stars become super-Eddington \citep{ekstrom2012grids} and contributes to a large fraction of the predicted mass loss. For comparison, at each time step we recompute a value for mass-loss rate using our \mdot-prescription. In this case, we have not measured values of \mdot\ below $\sim$10$^{-7}\msun\ yr^{-1}$ and so we regard this section of the plot as uncertain, although the contribution to overall mass lost in this region is negligible. Figure \ref{fig:mdot_geneva} shows \mdot-prescription being implemented in the Geneva stellar models is dependent only on the current luminosity of the star, leading to an over-prediction of the total mass lost during the RSG phase by up to a factor of 20. This result suggests stellar models could be over-predicting the number of stars that evolve to the blue of the HR, and hence {\it under-predict} the H-rich SN rate.  

\begin{figure}
	\centering
  	\caption{Change in current mass of 12, 15, 20 and 25\msun\ stars as a function of time.  }
  	\centering
  	\label{fig:mtot_geneva}
			    \includegraphics[width=\columnwidth]{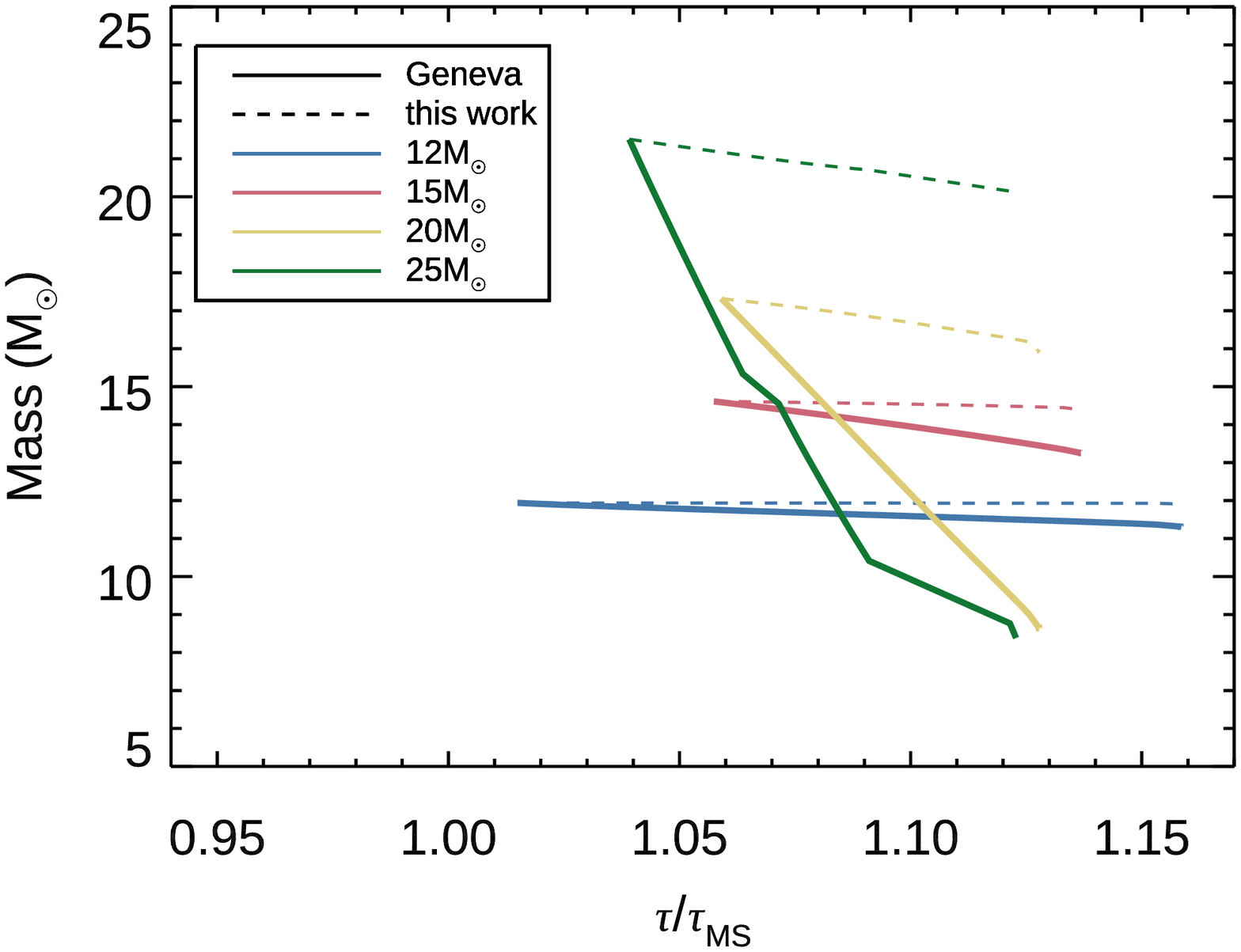}

\end{figure}

We now compare the predicted total amount of mass lost during the RSG phase ($M_{\rm tot}$) from the Geneva models and the \mdot\ prescription presented in this work, under the assumption that changing \mdot\ does not change the core evolution (and hence luminosity evolution) of the star. At each timestep a value for \mdot\ is calculated using the \lbol\ and initial mass of the star. Figure \ref{fig:mtot_geneva} shows the mass of the star as a function of time (scaled by MS lifetime). The solid lines show the mass of the star directly taken from the Geneva mass tracks (i.e. dJ88) and the dashed lines show the results when using the new \mdot-prescription. The current \mdot\ implementation in the Geneva models predicts a higher $M_{\rm tot}$ for all initial masses included here. Indeed, for the 20\msun\ star, we predict a total mass loss through the RSG phase of 1.4\msun\ while the current \mdot\ implementation in \citet{ekstrom2012grids} predicts a total mass loss of 9\msun. 
This is a dramatic difference.  With this new prescription, steady mass loss will be insufficient to allow single stars of 20-25 $M_{\odot}$ to evolve blueward to become LBVs, BSGs, or WR stars before exploding (see below).

\begin{figure}
	\centering
  	\caption{Total mass lost during the RSG phase compared to the mass of the envelope as a function of time.  }
  	\centering
  	\label{fig:menv}
			    \includegraphics[width=\columnwidth]{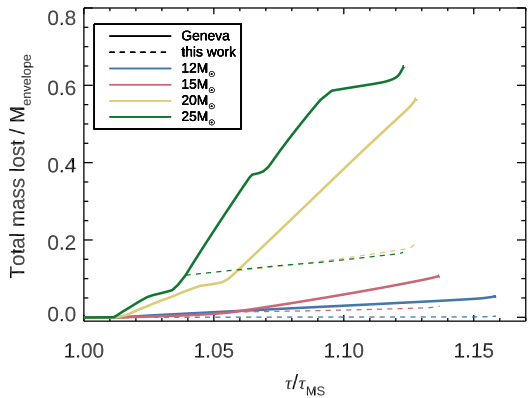}

\end{figure}

The factor which determines what kind of SN will be seen is the mass of the remaining H-rich envelope at core-collapse, and the density of the wind shortly before death. For stars which retain their envelope, the resulting SN will appear as a Type II H-rich SN, while those that lose their envelope will evolve to become WR (or BSG) stars before exploding as Type Ibc `stripped` SN. The Geneva mass tracks do not provide a value for envelope mass ($M_{\rm env}$) explicitly, and so we derive a lower limit for $M_{\rm env}$ by subtracting the convective core mass at the end of the MS from the mass of the star. We now use this to estimate the mass lost as a fraction of the envelope mass prior to SN, see Fig. \ref{fig:menv}. In this figure, the MS is plotted. The point at which the dashed line becomes visible is the point at which the RSG mass loss comes into effect. Our \mdot-prescription suggests that very little of the envelope mass is lost in the RSG phase, whereas the \mdot\ currently implemented in the Geneva models suggests as much as 50\% of the envelope can be lost during this period. It is this artificial loss of envelope mass that drives the stars back to the blue of the HR diagram \citep[see the 25\msun\ track in ][]{ekstrom2012grids}.  

The results of this study suggest that quiescent mass-loss during the RSG phase cannot be the sole evolutionary driver for massive stars. From the clusters studied here, there is no evidence for enhanced \mdot\ during the RSG phase and there is no physical motivation for stellar evolutionary models to ramp up \mdot\ in order to explain the RSG problem \cite[e.g.][]{georgy2013grids}, or to produce WR stars from single-stars via RSG mass loss.

\subsection{Caveats and assumptions}\label{section:caveats}
The work presented here represents a substantial improvement on dJ88, not just in the larger sample and homogeneous methodology, but also in the fact that we are able to isolate stellar mass as an independent variable. The previous attempt to  incorporate mass in the \mdot\ recipe \citep{nieuwenhuijzen1990parametrization} depended upon inferring mass from luminosities of field stars with uncertain distances and comparing to evolutionary tracks, where there is no unique mass-luminosity relation. Our strategy of using RSGs in stellar clusters, where the masses of the RSGs can be inferred from the age of the host cluster, has for the first time shown that high-mass RSGs are strongly discrepant from the dJ88 \mdot\ law.

Nevertheless, we have had to make some assumptions in the course of this work, which mean that we must provide caveats to our conclusions. We discuss these below.

\begin{itemize}
    \item Firstly, throughout this work we model the dusty shell around the RSGs as being spherically symmetric. It is well known that in reality the dust around RSGs is clumped and highly asymmetric \citep[e.g.][]{smith2001asymmetric,scicluna2015large}. This is an effect we studied in detail in \citet{beasor2016evolution}, where we found that even increasing the clumping to a filling factor of 50 has little-to-no effect on the output \mdot\ value provided the dust is optically thin (which is the case for all the RSGs included in our \mdot-prescription). Therefore, we concluded that clumping is unlikely to affect our \mdot\ measurements. However, in the case of a very dense wind, this could affect our \lbol\ measurements \citep[see the case of WOH G64 in][]{davies2018humphreys}. A dense wind could shift \lbol\ either higher or lower, depending on the orientation of the clumps. For example, when modelling the SED of WOH G64 many authors have noted that it cannot have a spherically symmetric dust shell \citep[see discussion in][]{davies2018humphreys}. In \citet{davies2018humphreys} the luminosity of WOH G64 was determined to be \logl$\sim$ 5.77 from integration under the SED. However, if the excess mid-IR emission originates in a dusty torus as suggested by \citet{ohnaka2008spatially}, the luminosity could be as low as \logl = 5.45. Given that this effect can shift stars either left or right in the \lbol-\mdot\ plot, it would introduce scatter into the relation rather than a systematic shift, so this is unlikely to affect our results.    

    \item We have assumed a uniform gas-to-dust ratio ($r_{\rm gd}$) for each of the RSGs, regardless of how close they are to supernova (for Galactic clusters we assume $r_{\rm gd}$ of 200, while for LMC clusters we assume $r_{\rm gd}$ of 500, see Section \ref{section:sedmodeling}). As the stars evolve towards SN and the dust shell mass increases, it is likely that the $r_{\rm gd}$ will change.  An $r_{\rm gd}$ that decreases with time (as the star makes dust more efficiently) would cause the slopes of the clusters in Fig. \ref{fig:mdotcompc4} to be less steep. This is something that could be checked with independent \mdot\ measurements derived from the gas, \citep[e.g. CO emission lines in the sub-mm,][]{decin2006probing}. However, note that such measurements are subject to their own uncertain correction factors, particularly the ratio of CO to H$_{2}$. One of our most important conclusions here is that the higher mass stars in RSGC1 have \mdot s that are strongly discrepant from dJ88. This conclusion is only undermined if the RSGs in RSGC1 have $r_{\rm gd}$ values that are significantly higher than 200. Indeed, for the mass-loss rates to be brought in line with the dJ88 prescription, all stars would require an $r_{\rm gd}$ value of $\sim$2500, which would be strongly discrepant with any previous measurement \citep[e.g.][]{goldman2017wind}.
    \item  From the typical RSG lifetime $T$ and the number of RSGs $N$ we have included we can state that we are likely to miss very extreme (i.e. rare) phases, but that these cannot last longer than $t \sim T/N$ and no more than $n\sim\sqrt{N}$ would have been missed. As any extreme mass-loss phase (\mdot\ > 10$^{-4}$\msun$yr^{-1}$) is likely to be very short ($\sim 10^4$yrs), the contribution of this to the total mass lost will only be on the order of $\sim$1\msun\ and hence will not have a significant effect on the onward evolution.  This can, of course, be simulated properly using an \mdot-$t$ distribution. 
    \item Finally, an important point to note is the sampling of RSG initial masses. We currently have four data points in our study, but unfortunately 3 of them have very similar ages. Since we have poor sampling in age between the youngest and oldest clusters, we simply used linear interpolation to estimate mass-loss rates for RSGs with intermediate masses. Adding further clusters with ages in the range 10--20Myr, rich in RSGs, would improve the precision of our work. Despite the poor sampling in age, the unequivocal result of this work is that the dJ88 mass-loss recipe grossly overestimates \mdot\ for the high-mass stars in RSGC1. The most massive RSGs are the objects that from a stellar evolution standpoint are the most interesting, as it is these stars that have previously been thought to have the strongest winds with the potential of stripping their own envelopes prior to explosion. It is the mass-loss of RSGs above 17\msun\ which is pertinent to understanding the upper mass cutoff for Type IIP SN progenitors \citep[the `Red Supergiant Problem'][]{smartt2009death,smartt2015observational} and the H-D limit \citep{humphreys1979studies,davies2018humphreys}. 

\end{itemize}

\subsection{Implications}
\subsubsection{LBVs as post-RSG objects}

Another evolutionary stage during which massive stars can lose a considerable amount of H-envelope mass is the luminous blue variable (LBV) phase. LBVs are hot massive stars, which exhibit large variations in brightness and powerful episodic mass-loss events. It was thought for a long time that all massive stars experience a brief LBV phase (10$^{4}$yr) prior to becoming Wolf-Rayet (WR) stars, where the strong episodic mass-loss can remove the majority of the remaining H-rich envelope \citep{humphreys1994luminous}. In this scenario, LBVs could not be the immediate progenitors of SNe, because they are followed by a WR phase.  However, more recent work has shown that LBVs are remarkably isolated from clusters of massive O-type stars \citep{st15}.  Their isolation requires that they have longer lifetimes than O stars or WR stars, making it impossible for LBVs to accomplish their presumed role in single-star evolution of removing the H envelope to transform O stars into WR stars. It was instead suggested that LBVs are likely products of binary evolution, whereby the LBV was initially a lower-mass star, but became more luminous because it is the mass-gainer of a system or the product of a merger \citep{st15,smith16,mojgan17}.  \citet{justham14} have modeled LBVs as merger products to explain how they can be potential SN progenitors.

Lower-luminosity LBVs might also be observed as SN progenitors if the LBV phase comes after the RSG phase.  \citet{groh2013massive} presented stellar evolution models that were coupled with radiative transfer modeling using CoMoving Frame GENeral \citep[CMFGEN,][]{hillier2001cmfgen} to predict the appearance of single SN progenitors prior to explosion. Using the \citeauthor{de1988mass} prescription for the RSG phase of evolution, the authors found that the 20 and 25\msun\ pre-SN spectra of the progenitors looked remarkably similar to those of LBVs, implying previously unknown evolutionary paths for lower-mass stars, 

$20\msun: RSG \longrightarrow BSG \longrightarrow LBV \longrightarrow SN;$

$25\msun:  RSG \longrightarrow WR \longrightarrow LBV \longrightarrow SN;$

\noindent Under this paradigm, following the RSG phase the star has shed enough mass to move back to the blue of the HR diagram and become a blue supergiant (BSG) before exploding as an LBV. A very similar path was predicted for the 25\msun\ model, but a WR, specifically an Ofpe/WN9-like, phase instead of a BSG phase. 

The LBV as post RSGs scenario is only viable if the mass-loss during the RSG phase is enough to evolve the star back to the blue. At present, we find the implementation of mass-loss in the Geneva models severely over-predicts the total amount of envelope mass lost during the RSG phase. This is exacerbated by the increase of \mdot\ by a further factor of 3 beyond the extrapolated prescription \citep{ekstrom2012grids,georgy2013grids} for stars with initial masses of 20\msun\ and above. The results of this paper show clearly that for stars of 20\msun\ and below, mass-loss during the RSG phase is not enough to remove the H-envelope and cause blue-ward motion. Though admittedly the empirical range of this study is 8-25\msun, it is unlikely that the envelope of 20--30\msun\ can be removed by quiescent RSG winds unless there is a large step change in \mdot\ for more massive RSGs. 

\subsubsection{SN interaction with CSM}
When SN~II-P progenitors explode there is observational evidence showing they crash into a dense CSM, such as their early light curves \citep[e.g.][]{morozova2017unifying}, and brief IIn phase that is observed in some SNe II-P \citep[e.g.][]{smith2015ptf11iqb,khazov2016flash}. In order to reproduce these observations, it has been claimed that the CSM must be very close to the star (i.e. within about 10au, or even within a stellar radius in some cases) and very dense \citep[e.g.][]{smith2015ptf11iqb,morozova2017unifying}. The light curves were modeled by \citet{moriya2018type} and \citep{morozova2017unifying}, who suggested \mdot\ values of 10$^{-3}$--10$^{-2}$\msun\ yr$^{-1}$. These mass-loss rates are substantially higher than we find for any object in our sample, although their cumulative mass lost is negligible because they only operate for a brief time shortly before the SN. In this work, we find that the amount of mass lost throughout the RSG phase (lasting approximately 10$^6$yrs) is very small even for the most massive progenitors. For a 25\msun\ RSG we predict only a total mass lost of 1.4\msun, which would correspond to approximately 8$\times$10$^{-4}$\msun\ of material within 1 stellar radius. A level of mass loss this low is unlikely to have an effect on the observed SN light curve \citep[e.g.][]{smith2017endurance}. Of course, this does not take into account any mass lost during potential periods of enhanced pre-SN mass-loss (see later), because our target stars used to derive the new prescription are not able to sample immediate SN progenitors (none of them have exploded yet).

To explain the apparent disagreement between the \mdot\ values found here and those being claimed for SN progenitors, we will now explore the methodology of \citet{moriya2018type} in more detail. The authors modeled RSGs with an acceleration zone to explain the rise times of several Type II-P SN light curves. By adopting wind acceleration parameter ($\beta$) values between 1-5, the authors conclude that the slow acceleration of the wind results in a dense CSM lying in the vicinity of the progenitor star upon explosion. However, as the $\beta$-law describes wind acceleration for radiatively driven winds \citep{castor1975radiation}, it is unclear if there is any justification in applying this to RSGs which likely have a very different driving mechanism. Though \citet{moriya2018type} study slowly accelerating winds ($\beta$=5), even this is likely far too fast for RSGs, where wind accelerates so slowly that the CSM is likely almost static within the first couple of stellar radii \citep{harper2001spatially,dessart2017explosion}.

Though we have shown that quiescent mass-loss is extremely ineffective at removing the envelope, we have not yet discussed how the envelope may be removed by a brief period of enhanced mass-loss, e.g. via binary envelope stripping or a short phase of enhanced mass-loss in the decades or centuries before explosion \citep[e.g.][]{sa14,smith2014mass,yc10,qs12}. \citet{davies2018humphreys} estimated how long a period of enhanced mass-loss would need to last to remove a large fraction of the hydrogen envelope. Assuming any star undergoing this enhanced \mdot\ would be visible as a maser emitter, \citet{davies2018humphreys} found four OH/IR emitters in their total sample of 73 RSGs with \logl\ $>$ 5. Assuming the RSG phase is $\sim$10$^{6}$yrs and that the superwind phase is a ubiquitous feature of single star evolution (which is by no means certain), this suggests any `superwind' phase is on the order of 10$^{4}$yrs. If the \mdot\ during this time is as high as that of the maser emitters in the \citet{goldman2017wind} sample ($\sim10^{-4}$\msun yr$^{-1}$), several Solar masses of envelope could potentially be lost. 
  
	
\section{Conclusions}
Mass-loss rate prescriptions must be assumed in stellar evolutionary codes to determine the fate of massive stars. While an \mdot\ relation found from first principles cannot be attained, models input empirically derived \mdot\ recipes. By using RSGs in clusters with known ages and initial masses, we derive a new mass-dependent mass-loss rate prescription that yields mass-loss rates lower than previous prescriptions used in stellar evolution models, and much lower than the artificially elevated mass-loss rates that are sometimes adopted. Below we outline the main conclusions of this work:

\begin{enumerate}
    \item There is no observationally motivated reason to increase the quiescent mass-loss rates of RSGs by factors of three or more above the dJ88 rate, as is currently implemented in the Geneva models. Indeed, we show the dJ88 rate is already a factor of 9 too high for the quiescent winds of massive RSGs. RSGs that evolve as single stars {\it cannot} shed their H-envelope through quiescent winds, and thus will die with this envelope intact (producing a Type II SN) in the absence of another stripping mechanism.
    \item Mass-loss rates derived from dust enshrouded stars 
    {\it should not} be used for quiescent RSG winds, as they are systematically too high by orders of magnitude for the majority of stars in the RSG phase. The dust enshrouded RSGs either represent a very small fraction of the RSG lifetime of a single star ($\sim 10^4$yrs), or are the product of another evolutionary channel (e.g. binary system, common envelope merger, mass gainer).
    \item Single stars with initial masses <25\msun\ do not lose enough mass through their quiescent winds to evolve blueward, and hence cannot create WR, BSG or LBV stars as some evolutionary models have predicted.
    
    \item If the \mdot-prescription derived here were implemented into stellar evolution models, stars with initial masses well in excess of 30\msun\ would fail to evolve back to the blue after becoming an RSG, leading to an upper luminosity limit (otherwise known as the Humphreys-Davidson limit) as high as \logl=6. This is in contrast with observations which show a clear cutoff at 5.5 \citep{davies2018humphreys}, implying quiescent RSG winds are not responsible for the upper luminosity limit. 
    \item The relative number of stripped/unstripped SN events predicted by single star stellar evolution models is likely incorrect, with the number of H-rich SN being underpredicted. However, this ratio could be heavily dominated by binary systems. 
\end{enumerate}


Our work here suggests that in contrast to what is predicted by single star evolutionary models, quiescent mass-loss during the RSG phase has little or no effect in stripping the envelope prior to SN. 


\section*{Acknowledgements}

We would like to thank the anonymous referee for useful comments which helped improve the paper. Based in part on observations made with the NASA/DLR Stratospheric Observatory for Infrared Astronomy (SOFIA). SOFIA is jointly operated by the Universities Space Research Association, Inc. (USRA), under NASA contract NNA17BF53C, and the Deutsches SOFIA Institut (DSI) under DLR contract 50 OK 0901 to the University of Stuttgart. Financial support for this work was provided by NASA through award \# 05 0064 issued by USRA. Support for this work was provided by NASA through Hubble Fellowship grant HST-HF2-51428 awarded by the Space Telescope Science Institute, which is operated by the Association of Universities for Research in Astronomy, Inc., for NASA, under contract NAS5-26555. RDG was supported by NASA and the United States Air Force.




\bibliographystyle{mnras}
\bibliography{references} 

\begin{thebibliography}{}
\makeatletter
\relax
\def\mn@urlcharsother{\let\do\@makeother \do\$\do\&\do\#\do\^\do\_\do\%\do\~}
\def\mn@doi{\begingroup\mn@urlcharsother \@ifnextchar [ {\mn@doi@}
  {\mn@doi@[]}}
\def\mn@doi@[#1]#2{\def\@tempa{#1}\ifx\@tempa\@empty \href
  {http://dx.doi.org/#2} {doi:#2}\else \href {http://dx.doi.org/#2} {#1}\fi
  \endgroup}
\def\mn@eprint#1#2{\mn@eprint@#1:#2::\@nil}
\def\mn@eprint@arXiv#1{\href {http://arxiv.org/abs/#1} {{\tt arXiv:#1}}}
\def\mn@eprint@dblp#1{\href {http://dblp.uni-trier.de/rec/bibtex/#1.xml}
  {dblp:#1}}
\def\mn@eprint@#1:#2:#3:#4\@nil{\def\@tempa {#1}\def\@tempb {#2}\def\@tempc
  {#3}\ifx \@tempc \@empty \let \@tempc \@tempb \let \@tempb \@tempa \fi \ifx
  \@tempb \@empty \def\@tempb {arXiv}\fi \@ifundefined
  {mn@eprint@\@tempb}{\@tempb:\@tempc}{\expandafter \expandafter \csname
  mn@eprint@\@tempb\endcsname \expandafter{\@tempc}}}

\bibitem[\protect\citeauthoryear{{Aghakhanloo}, {Murphy}, {Smith}  \&
  {Hlo{\v{z}}ek}}{{Aghakhanloo} et~al.}{2017}]{mojgan17}
{Aghakhanloo} M.,  {Murphy} J.~W.,  {Smith} N.,   {Hlo{\v{z}}ek} R.,  2017,
  \mn@doi [\mnras] {10.1093/mnras/stx2050}, \href
  {https://ui.adsabs.harvard.edu/abs/2017MNRAS.472..591A} {472, 591}

\bibitem[\protect\citeauthoryear{Beasor \& Davies}{Beasor \&
  Davies}{2016}]{beasor2016evolution}
Beasor E.~R.,  Davies B.,  2016, Monthly Notices of the Royal Astronomical
  Society, 463, 1269

\bibitem[\protect\citeauthoryear{Beasor \& Davies}{Beasor \&
  Davies}{2018}]{beasor2018evolution}
Beasor E.~R.,  Davies B.,  2018, Monthly Notices of the Royal Astronomical
  Society, 475, 55

\bibitem[\protect\citeauthoryear{{Beasor}, {Davies}, {Smith}  \&
  {Bastian}}{{Beasor} et~al.}{2019}]{beasor2019ages}
{Beasor} E.~R.,  {Davies} B.,  {Smith} N.,   {Bastian} N.,  2019, arXiv
  e-prints, \href {http://adsabs.harvard.edu/abs/2019arXiv190305106B} {}

\bibitem[\protect\citeauthoryear{Bressan, Marigo, Girardi, Salasnich, Dal~Cero,
  Rubele  \& Nanni}{Bressan et~al.}{2012}]{bressan2012parsec}
Bressan A.,  Marigo P.,  Girardi L.,  Salasnich B.,  Dal~Cero C.,  Rubele S.,
  Nanni A.,  2012, Monthly Notices of the Royal Astronomical Society, 427, 127

\bibitem[\protect\citeauthoryear{Brott et~al.,}{Brott
  et~al.}{2011}]{brott2011rotating}
Brott I.,  et~al., 2011, Astronomy \& Astrophysics, 530, A115

\bibitem[\protect\citeauthoryear{{Castor}, {Abbott}  \& {Klein}}{{Castor}
  et~al.}{1975}]{castor1975radiation}
{Castor} J.~I.,  {Abbott} D.~C.,   {Klein} R.~I.,  1975, \mn@doi [\apj]
  {10.1086/153315}, \href {http://adsabs.harvard.edu/abs/1975ApJ...195..157C}
  {195, 157}

\bibitem[\protect\citeauthoryear{{Choi}, {Dotter}, {Conroy}, {Cantiello},
  {Paxton}  \& {Johnson}}{{Choi} et~al.}{2016}]{choi2016mist}
{Choi} J.,  {Dotter} A.,  {Conroy} C.,  {Cantiello} M.,  {Paxton} B.,
  {Johnson} B.~D.,  2016, \mn@doi [\apj] {10.3847/0004-637X/823/2/102}, \href
  {http://adsabs.harvard.edu/abs/2016ApJ...823..102C} {823, 102}

\bibitem[\protect\citeauthoryear{Chugai et~al.,}{Chugai
  et~al.}{2004}]{chugai2004type}
Chugai N.~N.,  et~al., 2004, Monthly Notices of the Royal Astronomical Society,
  352, 1213

\bibitem[\protect\citeauthoryear{Cioni et~al.,}{Cioni
  et~al.}{2000}]{cioni2000denis}
Cioni M.-R.,  et~al., 2000, Astronomy and Astrophysics Supplement Series, 144,
  235

\bibitem[\protect\citeauthoryear{Currie et~al.,}{Currie
  et~al.}{2010}]{currie2010stellar}
Currie T.,  et~al., 2010, The Astrophysical Journal Supplement Series, 186, 191

\bibitem[\protect\citeauthoryear{Davies \& Beasor}{Davies \&
  Beasor}{2018}]{davies2018initial}
Davies B.,  Beasor E.~R.,  2018, Monthly Notices of the Royal Astronomical
  Society, 474, 2116

\bibitem[\protect\citeauthoryear{{Davies} \& {Beasor}}{{Davies} \&
  {Beasor}}{2019}]{davies2019distances}
{Davies} B.,  {Beasor} E.,  2019, arXiv e-prints, \href
  {http://adsabs.harvard.edu/abs/2019arXiv190312506D} {}

\bibitem[\protect\citeauthoryear{Davies, Figer, Law, Kudritzki, Najarro,
  Herrero  \& MacKenty}{Davies et~al.}{2008}]{davies2008cool}
Davies B.,  Figer D.~F.,  Law C.~J.,  Kudritzki R.-P.,  Najarro F.,  Herrero
  A.,   MacKenty J.~W.,  2008, The Astrophysical Journal, 676, 1016

\bibitem[\protect\citeauthoryear{Davies et~al.,}{Davies
  et~al.}{2013}]{davies2013temperatures}
Davies B.,  et~al., 2013, The Astrophysical Journal, 767, 3

\bibitem[\protect\citeauthoryear{Davies, Crowther  \& Beasor}{Davies
  et~al.}{2018}]{davies2018humphreys}
Davies B.,  Crowther P.~A.,   Beasor E.~R.,  2018, Monthly Notices of the Royal
  Astronomical Society, 478, 3138

\bibitem[\protect\citeauthoryear{{Decin}, {Hony}, {de Koter}, {Justtanont},
  {Tielens}  \& {Waters}}{{Decin} et~al.}{2006}]{decin2006probing}
{Decin} L.,  {Hony} S.,  {de Koter} A.,  {Justtanont} K.,  {Tielens}
  A.~G.~G.~M.,   {Waters} L.~B.~F.~M.,  2006, \mn@doi [\aap]
  {10.1051/0004-6361:20065230}, \href
  {http://adsabs.harvard.edu/abs/2006A%26A...456..549D} {456, 549}

\bibitem[\protect\citeauthoryear{{Dessart}, {John Hillier}  \&
  {Audit}}{{Dessart} et~al.}{2017}]{dessart2017explosion}
{Dessart} L.,  {John Hillier} D.,   {Audit} E.,  2017, \mn@doi [\aap]
  {10.1051/0004-6361/201730942}, \href
  {https://ui.adsabs.harvard.edu/abs/2017A&A...605A..83D} {605, A83}

\bibitem[\protect\citeauthoryear{Doggett \& Branch}{Doggett \&
  Branch}{1985}]{doggett1985comparative}
Doggett J.,  Branch D.,  1985, The Astronomical Journal, 90, 2303

\bibitem[\protect\citeauthoryear{Dotter}{Dotter}{2016}]{dotter2016mesa}
Dotter A.,  2016, The Astrophysical Journal Supplement Series, 222, 8

\bibitem[\protect\citeauthoryear{Draine \& Lee}{Draine \&
  Lee}{1984}]{draine1984optical}
Draine B.,  Lee H.~M.,  1984, The Astrophysical Journal, 285, 89

\bibitem[\protect\citeauthoryear{Ekstr{\"o}m et~al.,}{Ekstr{\"o}m
  et~al.}{2012}]{ekstrom2012grids}
Ekstr{\"o}m S.,  et~al., 2012, Astronomy \& Astrophysics, 537, A146

\bibitem[\protect\citeauthoryear{Epchtein et~al.,}{Epchtein
  et~al.}{1994}]{epchtein1994denis}
Epchtein N.,  et~al., 1994, in , Science with Astronomical Near-Infrared Sky
  Surveys.
Springer, pp~3--9

\bibitem[\protect\citeauthoryear{Fazio et~al.,}{Fazio
  et~al.}{2004}]{fazio2004irac}
Fazio G.,  et~al., 2004, The Astrophysical Journal Supplement Series, 154, 10

\bibitem[\protect\citeauthoryear{Figer, MacKenty, Robberto, Smith, Najarro,
  Kudritzki  \& Herrero}{Figer et~al.}{2006}]{figer2006discovery}
Figer D.~F.,  MacKenty J.~W.,  Robberto M.,  Smith K.,  Najarro F.,  Kudritzki
  R.~P.,   Herrero A.,  2006, The Astrophysical Journal, 643, 1166

\bibitem[\protect\citeauthoryear{Gaia~Collaboration et~al.,}{Gaia~Collaboration
  et~al.}{2018}]{gaiadr2}
Gaia~Collaboration B.~A.,  et~al., 2018, Astronomy \& astrophysics, 616, A1

\bibitem[\protect\citeauthoryear{Gehrz et~al.,}{Gehrz
  et~al.}{2007}]{gehrz2007spitzer}
Gehrz R.~D.,  et~al., 2007, Review of scientific instruments, 78, 011302

\bibitem[\protect\citeauthoryear{Gehrz, Becklin, De~Pater, Lester, Roellig  \&
  Woodward}{Gehrz et~al.}{2009}]{gehrz2009new}
Gehrz R.~D.,  Becklin E.,  De~Pater I.,  Lester D.,  Roellig T.,   Woodward
  C.~E.,  2009, Advances in Space Research, 44, 413

\bibitem[\protect\citeauthoryear{Georgy}{Georgy}{2012}]{georgy2012yellow}
Georgy C.,  2012, Astronomy \& Astrophysics, 538, L8

\bibitem[\protect\citeauthoryear{Georgy et~al.,}{Georgy
  et~al.}{2013}]{georgy2013grids}
Georgy C.,  et~al., 2013, Astronomy \& Astrophysics, 558, A103

\bibitem[\protect\citeauthoryear{Goldman et~al.,}{Goldman
  et~al.}{2017}]{goldman2017wind}
Goldman S.~R.,  et~al., 2017, Monthly Notices of the Royal Astronomical
  Society, 465, 403

\bibitem[\protect\citeauthoryear{{Groh}, {Meynet}  \& {Ekstr{\"o}m}}{{Groh}
  et~al.}{2013}]{groh2013massive}
{Groh} J.~H.,  {Meynet} G.,   {Ekstr{\"o}m} S.,  2013, \mn@doi [\aap]
  {10.1051/0004-6361/201220741}, \href
  {http://adsabs.harvard.edu/abs/2013A%26A...550L...7G} {550, L7}

\bibitem[\protect\citeauthoryear{Gustafsson, Edvardsson, Eriksson,
  J{\o}rgensen, Nordlund  \& Plez}{Gustafsson
  et~al.}{2008}]{gustafsson2008grid}
Gustafsson B.,  Edvardsson B.,  Eriksson K.,  J{\o}rgensen U.~G.,  Nordlund
  {\AA}.,   Plez B.,  2008, Astronomy \& Astrophysics, 486, 951

\bibitem[\protect\citeauthoryear{Harper, Brown  \& Lim}{Harper
  et~al.}{2001}]{harper2001spatially}
Harper G.~M.,  Brown A.,   Lim J.,  2001, The Astrophysical Journal, 551, 1073

\bibitem[\protect\citeauthoryear{Herter et~al.,}{Herter
  et~al.}{2012}]{herter2012first}
Herter T.,  et~al., 2012, The Astrophysical Journal Letters, 749, L18

\bibitem[\protect\citeauthoryear{{Hillier} \& {Lanz}}{{Hillier} \&
  {Lanz}}{2001}]{hillier2001cmfgen}
{Hillier} D.~J.,  {Lanz} T.,  2001, in {Ferland} G.,  {Savin} D.~W.,  eds,
  Astronomical Society of the Pacific Conference Series Vol. 247, Spectroscopic
  Challenges of Photoionized Plasmas. p.~343

\bibitem[\protect\citeauthoryear{Houck et~al.,}{Houck
  et~al.}{2004}]{houck2004irs}
Houck J.~R.,  et~al., 2004, The Astrophysical Journal Supplement Series, 154,
  18

\bibitem[\protect\citeauthoryear{Humphreys \& Davidson}{Humphreys \&
  Davidson}{1979}]{humphreys1979studies}
Humphreys R.~M.,  Davidson K.,  1979, The Astrophysical Journal, 232, 409

\bibitem[\protect\citeauthoryear{Humphreys \& Davidson}{Humphreys \&
  Davidson}{1994}]{humphreys1994luminous}
Humphreys R.~M.,  Davidson K.,  1994, Publications of the Astronomical Society
  of the Pacific, 106, 1025

\bibitem[\protect\citeauthoryear{Ivezic, Nenkova  \& Elitzur}{Ivezic
  et~al.}{1999}]{ivezic1999dusty}
Ivezic Z.,  Nenkova M.,   Elitzur M.,  1999, Astrophysics Source Code Library,
  1, 11001

\bibitem[\protect\citeauthoryear{Javadi, van Loon, Khosroshahi  \&
  Mirtorabi}{Javadi et~al.}{2013}]{javadi2013uk}
Javadi A.,  van Loon J.~T.,  Khosroshahi H.,   Mirtorabi M.~T.,  2013, Monthly
  Notices of the Royal Astronomical Society, 432, 2824

\bibitem[\protect\citeauthoryear{{Justham}, {Podsiadlowski}  \&
  {Vink}}{{Justham} et~al.}{2014}]{justham14}
{Justham} S.,  {Podsiadlowski} P.,   {Vink} J.~S.,  2014, \mn@doi [\apj]
  {10.1088/0004-637X/796/2/121}, \href
  {https://ui.adsabs.harvard.edu/abs/2014ApJ...796..121J} {796, 121}

\bibitem[\protect\citeauthoryear{Khazov et~al.,}{Khazov
  et~al.}{2016}]{khazov2016flash}
Khazov D.,  et~al., 2016, The Astrophysical Journal, 818, 3

\bibitem[\protect\citeauthoryear{Koornneef}{Koornneef}{1983}]{koornneef1983near}
Koornneef J.,  1983, Astronomy and Astrophysics, 128, 84

\bibitem[\protect\citeauthoryear{Levesque, Massey, Olsen, Plez, Josselin,
  Maeder  \& Meynet}{Levesque et~al.}{2005}]{levesque2005effective}
Levesque E.~M.,  Massey P.,  Olsen K.,  Plez B.,  Josselin E.,  Maeder A.,
  Meynet G.,  2005, The Astrophysical Journal, 628, 973

\bibitem[\protect\citeauthoryear{Marco \& Negueruela}{Marco \&
  Negueruela}{2013}]{marco2013ngc}
Marco A.,  Negueruela I.,  2013, Astronomy \& Astrophysics, 552, A92

\bibitem[\protect\citeauthoryear{Marshall, van Loon, Matsuura, Wood, Zijlstra
  \& Whitelock}{Marshall et~al.}{2004}]{marshall2004asymptotic}
Marshall J.~R.,  van Loon J.~T.,  Matsuura M.,  Wood P.~R.,  Zijlstra A.~A.,
  Whitelock P.~A.,  2004, Monthly Notices of the Royal Astronomical Society,
  355, 1348

\bibitem[\protect\citeauthoryear{{Massey} \& {Olsen}}{{Massey} \&
  {Olsen}}{2003}]{massey2003evolution}
{Massey} P.,  {Olsen} K.~A.~G.,  2003, \mn@doi [\aj] {10.1086/379558}, \href
  {http://adsabs.harvard.edu/abs/2003AJ....126.2867M} {126, 2867}

\bibitem[\protect\citeauthoryear{{Mathis}}{{Mathis}}{1990}]{Mathis90}
{Mathis} J.~S.,  1990, \mn@doi [\araa] {10.1146/annurev.aa.28.090190.000345},
  \href {https://ui.adsabs.harvard.edu/abs/1990ARA&A..28...37M} {28, 37}

\bibitem[\protect\citeauthoryear{Mauron \& Josselin}{Mauron \&
  Josselin}{2011}]{mauron2011mass}
Mauron N.,  Josselin E.,  2011, Astronomy \& Astrophysics, 526, A156

\bibitem[\protect\citeauthoryear{Meynet \& Maeder}{Meynet \&
  Maeder}{2003}]{meynet2003stellar}
Meynet G.,  Maeder A.,  2003, Astronomy \& Astrophysics, 404, 975

\bibitem[\protect\citeauthoryear{{Moriya}, {F{\"o}rster}, {Yoon},
  {Gr{\"a}fener}  \& {Blinnikov}}{{Moriya} et~al.}{2018}]{moriya2018type}
{Moriya} T.~J.,  {F{\"o}rster} F.,  {Yoon} S.-C.,  {Gr{\"a}fener} G.,
  {Blinnikov} S.~I.,  2018, \mn@doi [\mnras] {10.1093/mnras/sty475}, \href
  {http://adsabs.harvard.edu/abs/2018MNRAS.476.2840M} {476, 2840}

\bibitem[\protect\citeauthoryear{Morozova, Piro  \& Valenti}{Morozova
  et~al.}{2017}]{morozova2017unifying}
Morozova V.,  Piro A.~L.,   Valenti S.,  2017, The Astrophysical Journal, 838,
  28

\bibitem[\protect\citeauthoryear{Niederhofer, Hilker, Bastian  \&
  Silva-Villa}{Niederhofer et~al.}{2015}]{niederhofer2015no}
Niederhofer F.,  Hilker M.,  Bastian N.,   Silva-Villa E.,  2015, Astronomy \&
  Astrophysics, 575, A62

\bibitem[\protect\citeauthoryear{Nieuwenhuijzen \& de Jager}{Nieuwenhuijzen \&
  de~Jager}{1990}]{nieuwenhuijzen1990parametrization}
Nieuwenhuijzen H.,  de Jager C.,  1990, Astronomy and Astrophysics, 231, 134

\bibitem[\protect\citeauthoryear{Ohnaka, Driebe, Hofmann, Weigelt  \&
  Wittkowski}{Ohnaka et~al.}{2008}]{ohnaka2008spatially}
Ohnaka K.,  Driebe T.,  Hofmann K.-H.,  Weigelt G.,   Wittkowski M.,  2008,
  Astronomy \& Astrophysics, 484, 371

\bibitem[\protect\citeauthoryear{Paxton, Bildsten, Dotter, Herwig, Lesaffre  \&
  Timmes}{Paxton et~al.}{2010}]{paxton2010modules}
Paxton B.,  Bildsten L.,  Dotter A.,  Herwig F.,  Lesaffre P.,   Timmes F.,
  2010, The Astrophysical Journal Supplement Series, 192, 3

\bibitem[\protect\citeauthoryear{Paxton et~al.,}{Paxton
  et~al.}{2013}]{paxton2013modules}
Paxton B.,  et~al., 2013, The Astrophysical Journal Supplement Series, 208, 4

\bibitem[\protect\citeauthoryear{Paxton et~al.,}{Paxton
  et~al.}{2015}]{paxton2015modules}
Paxton B.,  et~al., 2015, The Astrophysical Journal Supplement Series, 220, 15

\bibitem[\protect\citeauthoryear{Pietrzy{\'n}ski et~al.,}{Pietrzy{\'n}ski
  et~al.}{2013}]{pietrzynski2013eclipsing}
Pietrzy{\'n}ski G.,  et~al., 2013, Nature, 495, 76

\bibitem[\protect\citeauthoryear{Plez}{Plez}{2012}]{plez2012turbospectrum}
Plez B.,  2012, Astrophysics Source Code Library, 1, 05004

\bibitem[\protect\citeauthoryear{{Quataert} \& {Shiode}}{{Quataert} \&
  {Shiode}}{2012}]{qs12}
{Quataert} E.,  {Shiode} J.,  2012, \mn@doi [\mnras]
  {10.1111/j.1745-3933.2012.01264.x}, \href
  {https://ui.adsabs.harvard.edu/abs/2012MNRAS.423L..92Q} {423, L92}

\bibitem[\protect\citeauthoryear{Richards \& Yates}{Richards \&
  Yates}{1998}]{richards1998maser}
Richards A.,  Yates J.,  1998, Irish Astronomical Journal, 25, 7

\bibitem[\protect\citeauthoryear{Scicluna, Siebenmorgen, Wesson, Blommaert,
  Kasper, Voshchinnikov  \& Wolf}{Scicluna et~al.}{2015}]{scicluna2015large}
Scicluna P.,  Siebenmorgen R.,  Wesson R.,  Blommaert J.,  Kasper M.,
  Voshchinnikov N.,   Wolf S.,  2015, Online Material p, 1

\bibitem[\protect\citeauthoryear{Skrutskie et~al.,}{Skrutskie
  et~al.}{2006}]{skrutskie2006two2}
Skrutskie M.,  et~al., 2006, The Astronomical Journal, 131, 1163

\bibitem[\protect\citeauthoryear{Smartt}{Smartt}{2015}]{smartt2015observational}
Smartt S.,  2015, Publications of the Astronomical Society of Australia, 32,
  e016

\bibitem[\protect\citeauthoryear{Smartt, Eldridge, Crockett  \& Maund}{Smartt
  et~al.}{2009}]{smartt2009death}
Smartt S.,  Eldridge J.,  Crockett R.,   Maund J.~R.,  2009, Monthly Notices of
  the Royal Astronomical Society, 395, 1409

\bibitem[\protect\citeauthoryear{Smith}{Smith}{2014}]{smith2014mass}
Smith N.,  2014, Annual Review of Astronomy and Astrophysics, 52, 487

\bibitem[\protect\citeauthoryear{{Smith}}{{Smith}}{2016}]{smith16}
{Smith} N.,  2016, \mn@doi [\mnras] {10.1093/mnras/stw1533}, \href
  {https://ui.adsabs.harvard.edu/abs/2016MNRAS.461.3353S} {461, 3353}

\bibitem[\protect\citeauthoryear{{Smith} \& {Arnett}}{{Smith} \&
  {Arnett}}{2014}]{sa14}
{Smith} N.,  {Arnett} W.~D.,  2014, \mn@doi [\apj]
  {10.1088/0004-637X/785/2/82}, \href
  {https://ui.adsabs.harvard.edu/abs/2014ApJ...785...82S} {785, 82}

\bibitem[\protect\citeauthoryear{{Smith} \& {Tombleson}}{{Smith} \&
  {Tombleson}}{2015}]{st15}
{Smith} N.,  {Tombleson} R.,  2015, \mn@doi [\mnras] {10.1093/mnras/stu2430},
  \href {https://ui.adsabs.harvard.edu/abs/2015MNRAS.447..598S} {447, 598}

\bibitem[\protect\citeauthoryear{Smith, Humphreys, Davidson, Gehrz, Schuster
  \& Krautter}{Smith et~al.}{2001}]{smith2001asymmetric}
Smith N.,  Humphreys R.~M.,  Davidson K.,  Gehrz R.~D.,  Schuster M.,
  Krautter J.,  2001, The Astronomical Journal, 121, 1111

\bibitem[\protect\citeauthoryear{{Smith}, {Hinkle}  \& {Ryde}}{{Smith}
  et~al.}{2009}]{smith09}
{Smith} N.,  {Hinkle} K.~H.,   {Ryde} N.,  2009, \mn@doi [\aj]
  {10.1088/0004-6256/137/3/3558}, \href
  {https://ui.adsabs.harvard.edu/abs/2009AJ....137.3558S} {137, 3558}

\bibitem[\protect\citeauthoryear{{Smith}, {Li}, {Filippenko}  \&
  {Chornock}}{{Smith} et~al.}{2011}]{smith11}
{Smith} N.,  {Li} W.,  {Filippenko} A.~V.,   {Chornock} R.,  2011, \mn@doi
  [\mnras] {10.1111/j.1365-2966.2011.17229.x}, \href
  {https://ui.adsabs.harvard.edu/abs/2011MNRAS.412.1522S} {412, 1522}

\bibitem[\protect\citeauthoryear{Smith et~al.,}{Smith
  et~al.}{2015}]{smith2015ptf11iqb}
Smith N.,  et~al., 2015, Monthly Notices of the Royal Astronomical Society,
  449, 1876

\bibitem[\protect\citeauthoryear{Smith et~al.,}{Smith
  et~al.}{2016}]{smith2017endurance}
Smith N.,  et~al., 2016, Monthly Notices of the Royal Astronomical Society, p.
  stw3204

\bibitem[\protect\citeauthoryear{{Wang} \& {Chen}}{{Wang} \&
  {Chen}}{2019}]{Wang-Chen19}
{Wang} S.,  {Chen} X.,  2019, \mn@doi [\apj] {10.3847/1538-4357/ab1c61}, \href
  {https://ui.adsabs.harvard.edu/abs/2019ApJ...877..116W} {877, 116}

\bibitem[\protect\citeauthoryear{Werner et~al.,}{Werner
  et~al.}{2004}]{werner2004spitzer}
Werner M.,  et~al., 2004, The Astrophysical Journal Supplement Series, 154, 1

\bibitem[\protect\citeauthoryear{Wright et~al.,}{Wright
  et~al.}{2010}]{wright2010wide}
Wright E.~L.,  et~al., 2010, The Astronomical Journal, 140, 1868

\bibitem[\protect\citeauthoryear{{Xue}, {Jiang}, {Gao}, {Liu}, {Wang}  \&
  {Li}}{{Xue} et~al.}{2016}]{Xue16}
{Xue} M.,  {Jiang} B.~W.,  {Gao} J.,  {Liu} J.,  {Wang} S.,   {Li} A.,  2016,
  \mn@doi [\apjs] {10.3847/0067-0049/224/2/23}, \href
  {https://ui.adsabs.harvard.edu/abs/2016ApJS..224...23X} {224, 23}

\bibitem[\protect\citeauthoryear{{Yoon} \& {Cantiello}}{{Yoon} \&
  {Cantiello}}{2010}]{yc10}
{Yoon} S.-C.,  {Cantiello} M.,  2010, \mn@doi [\apj]
  {10.1088/2041-8205/717/1/L62}, \href
  {https://ui.adsabs.harvard.edu/abs/2010ApJ...717L..62Y} {717, L62}

\bibitem[\protect\citeauthoryear{Young et~al.,}{Young
  et~al.}{2012}]{young2012early}
Young E.,  et~al., 2012, The Astrophysical Journal Letters, 749, L17

\bibitem[\protect\citeauthoryear{{Zaritsky}, {Harris}, {Thompson}  \&
  {Grebel}}{{Zaritsky} et~al.}{2004}]{zaritsky2004magellanic}
{Zaritsky} D.,  {Harris} J.,  {Thompson} I.~B.,   {Grebel} E.~K.,  2004,
  \mn@doi [\aj] {10.1086/423910}, \href
  {http://adsabs.harvard.edu/abs/2004AJ....128.1606Z} {128, 1606}

\bibitem[\protect\citeauthoryear{de Jager, Nieuwenhuijzen  \& Van
  Der~Hucht}{de~Jager et~al.}{1988}]{de1988mass}
de Jager C.,  Nieuwenhuijzen H.,   Van Der~Hucht K.,  1988, Astronomy and
  Astrophysics Supplement Series, 72, 259

\bibitem[\protect\citeauthoryear{{van Loon}}{{van Loon}}{2010}]{van2009effects}
{van Loon} J.~T.,  2010, in {Leitherer} C.,  {Bennett} P.~D.,  {Morris} P.~W.,
   {van Loon} J.~T.,  eds,  Astronomical Society of the Pacific Conference
  Series Vol. 425, Hot and Cool: Bridging Gaps in Massive Star Evolution.
  p.~279 (\mn@eprint {arXiv} {0906.4855})

\bibitem[\protect\citeauthoryear{van Loon, Zijlstra, Bujarrabal  \& Nyman}{van
  Loon et~al.}{2001}]{van2001circumstellar}
van Loon J.~T.,  Zijlstra A.~A.,  Bujarrabal V.,   Nyman L.-{\AA}.,  2001,
  Astronomy \& Astrophysics, 368, 950

\bibitem[\protect\citeauthoryear{van Loon, Cioni, Zijlstra  \& Loup}{van Loon
  et~al.}{2005}]{van2005empirical}
van Loon J.~T.,  Cioni M.-R.,  Zijlstra A.~A.,   Loup C.,  2005, Astronomy \&
  Astrophysics, 438, 273

\makeatother
\end{thebibliography}

\appendix

\section{Extinction law towards RSGC1}\label{section:appendixA}
The cluster RSGC1, located at the end of the Galactic Bar at a distance of 6.6kpc from Earth, is heavily obscurred. The extinction at 2$\mu$m is greater than 2mags \citep{davies2008cool}, and as such the extinction in the mid-IR is non-negligible. Measurements of the extinction law at mid-IR wavelengths are scarce, and seem to depend on sightline \citep{Mathis90,Xue16,Wang-Chen19}. Furthermore, the mid-IR extinction law is non-monotonic as it features absorption from silicate dust grains, which happens also to be the diagnostic feature we measure in emission to determine RSG mass-loss rates. It is therefore crucial for this work that we make an accurate measurement of the extinction law towards RSGC1.

Our methodology can be summarised as follows. We have obtained mid-IR spectroscopy of the RSGs in the cluster from the archives. We assume that the faintest RSG in RSGC1 (F14) has no intrinsic mid-IR excess. We then take the ratio of F14's spectrum to that of an appropriate model atmosphere to be a measurement of the extinction. The method is described in more detail below. 

\subsection{Benchmark object}
For our testbed object, we selected the RSG star F14. The star is bright enough to be easily detectable in the mid-IR, and comparatively spatially isolated, allowing us to combine reliable photometry and spectrophotometry to obtain a spectrum which is well flux-calibrated (see below). Whilst being bright, the indications are that the star has little or no infrared excess \citep{davies2008cool}. By comparing the star's spectrum to a model atmosphere, we can therefore determine the extinction as a function of wavelength. We note that, if the star {\it does} have some mid-IR excess, this would cause us to underestimate the extinction in the mid-IR relative to that in the near-IR. 

\subsection{Data}
For our mid-IR spectroscopy, we use the data from Spitzer/IRS program ID 40224 (PI B.\ Davies). The programme uses the low resolution mode, covering 5-15$\mu$m, and the high-resolution mode covering 10-35$\mu$m. The lo-res data has the advantage of a longer slit, which allows for accurate sky subtraction, and is known to provide excellent flux calibration. However, the wavelength coverage does not go to long enough wavelengths for our purposes. The hi-res data has spectral coverage which extends to longer wavelengths, however the shorter slit means that sky subtraction has to be done using dedicated sky observations, and the field around RSGC1 can be seen in mid-IR images to have patchy background emission. Furthermore, the flux calibration in hi-res IRS data is poor due to the slit covering only a fraction of the point spread function, making the whole dataset unreliable unless it can be independently flux calibrated. 

To provide flux calibration data for these spectra, we complement with mid-IR photometry. Though F14 is outside the field-of-view of our SOFIA data (presented here), the star is isolated enough to have reliable photometry in the lower spatial resolution images of MSX, as well as being covered by IRAC in the Spitzer/GLIMPSE survey. This means that we are able to reliably flux-calibrate the IRS spectra shortward of $\sim$20$\mu$m. Longer than 20$\mu$m, we are reliant on the Long-High (LH) IRS module, where the flux calibration is poor. To tune up the flux calibration at these longer wavelengths, we extract all IRS spectra for RSGC1 stars that are reasonably isolated (F6, F7, F10, F11, F13) and recalibrate the LH IRS data using the SOFIA photometry. This was achieved by applying a uniform scale factor of 0.62 to the LH spectra, which resulted in fluxes consistent with the long wavelength SOFIA spectra to within $\pm$5\%.

\begin{figure}
    \centering
    \includegraphics[width=8.5cm]{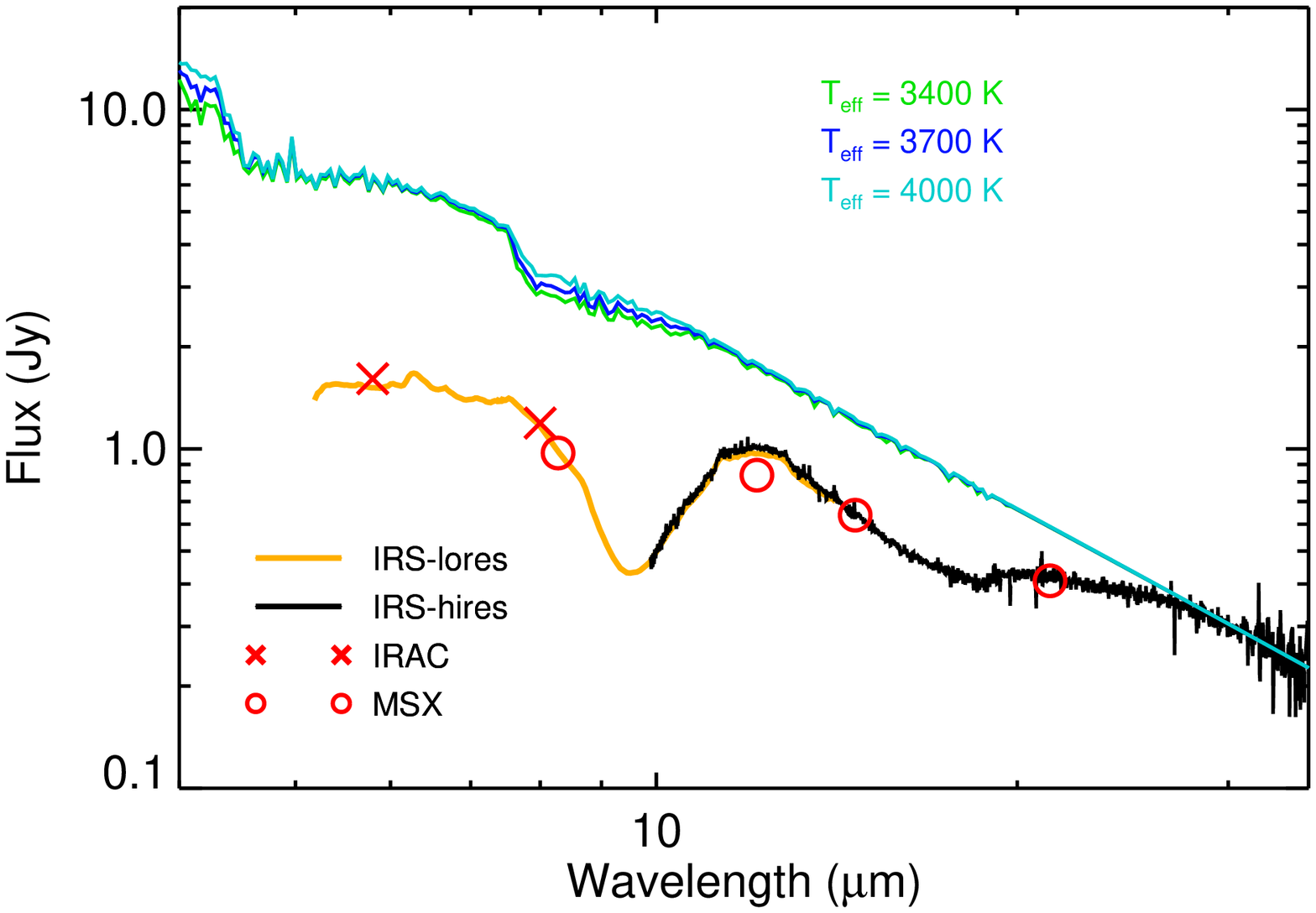}
    \caption{Mid-infrared spectroscopy and photometry of the star F14. Overplotted are spectra generated from MARCS model atmospheres at three different effective temperatures.}
    \label{fig:F14}
\end{figure}

In Fig.\ \ref{fig:F14} we plot the IRS spectra of F14, as well as the photometry from MSX and IRAC. The plot shows that, in the case of this star, there is excellent agreement between the spectroscopic and photometric data. 

\begin{figure} 
    \centering
    \includegraphics[width=8.5cm]{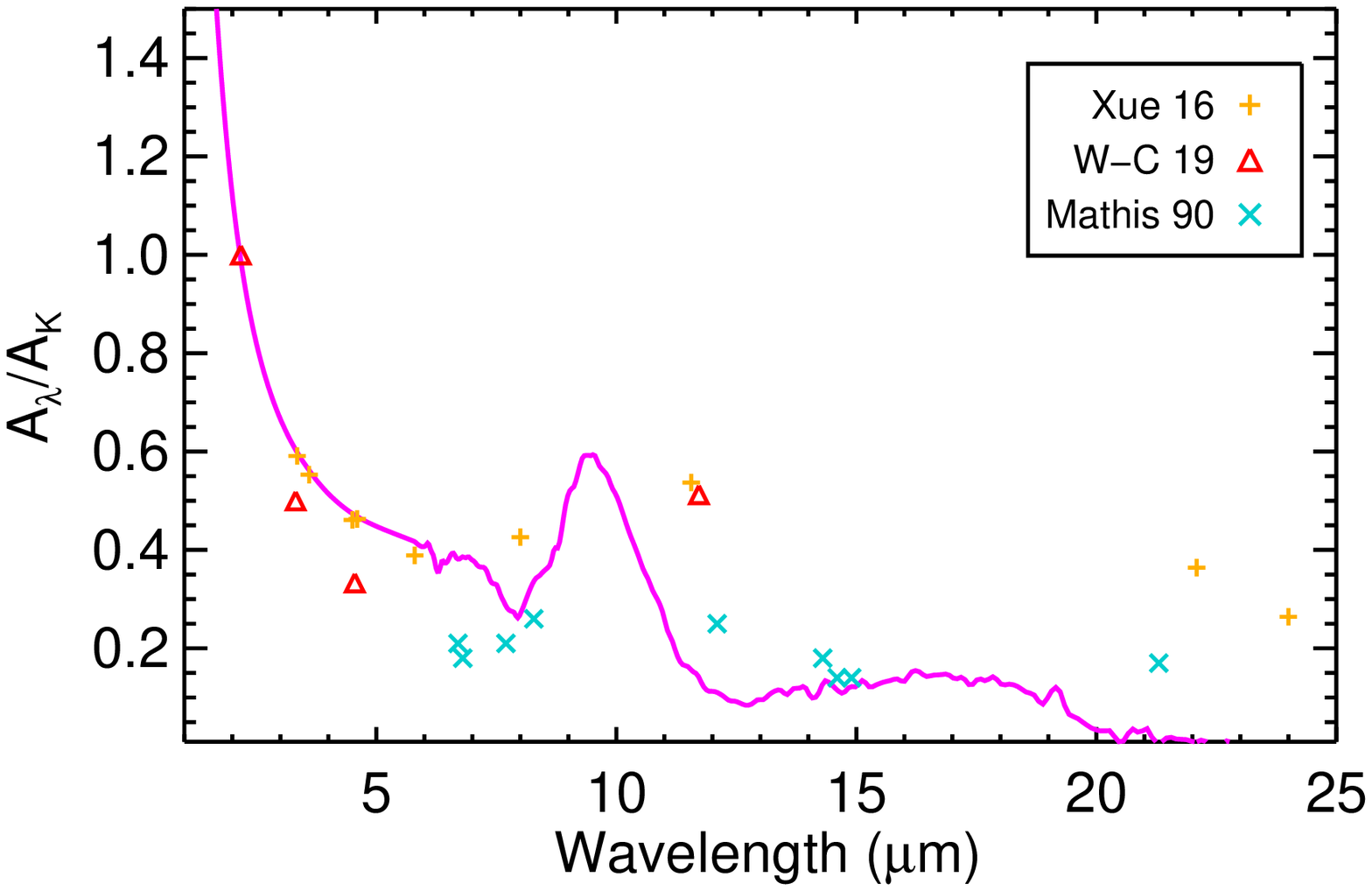}
    \caption{Extinction law from optical to mid-IR. Other measurements of the extinction law from the literature are overplotted as coloured points. }
    \label{fig:extlaw}
\end{figure}

\subsection{Determining the extinction law}
To infer the extinction law, we first require an estimate of the intrinsic spectrum of F14. For this purpose, we take MARCS model atmospheres \citep{gustafsson2008grid} with effective temperatures $T_{\rm}$ between 3400K and 4000K, gravity $\log(g/{\rm cgs} )= 0.0$, and Solar metallicity. The spectra are computed with the code TURBOSPECTRUM \citep{plez2012turbospectrum}. The model spectrum is then scaled to match the dereddened $K$-band photometry of F14, assuming an extinction of $A_{K} = 2.74$\footnote{This value of the extinction towards RSGC1 is taken from \citet{figer2006discovery}, where it was estimated from the average 2MASS colour excess of all RSGs in the cluster, assuming the average intrinsic colours of M supergiants. Though the extinction towards RSGC1 was updated in \citet{davies2008cool}, these authors estimated the extinction towards each individual star, assuming the intrinsic colours from spectral types estimated from $K$-band spectra. Since these spectral types are necessailty uncertain, we adopt the extinction estimated in Figer et al.}. Model spectra at three different values of $T_{\rm eff}$ are plotted over F14's IRS spectrum in Fig.\ \ref{fig:F14}. 

We estimate the extinction per unit wavelength by taking the ratio of the scaled model to the observed spectrum and applying the magnitude formula. The result is plotted as the magenta line in Fig.\ \ref{fig:extlaw}. Overplotted are the results from similar studies from the literature \citep{Mathis90,Xue16,Wang-Chen19}. Though the various studies serve to illustrate the uncertainties on the mid-IR extinction law and its dependence on sightline, the studies agree to within a factor of $\sim$2-3 at all wavelengths shorter than $\sim$20$\mu$m. Above 20$\mu$m, our extinction law falls to close to zero, whereas the other studies indicate that it remains roughly constant above $\sim$15$\mu$m. We caution that the location at which our results deviate from the other studies corresponds to the join between the Long-Low and Long-High IRS modules, and could be an artefact of poor flux calibration. To investigate the impact of any systematic error here, we experimented with two extinction laws: that shown in Fig.\ \ref{fig:extlaw}, and one that remains flat above 18$\mu$m. The mass-loss rates found using each of the extinction laws were consistent to within the errors. 

\label{lastpage}
\end{document}